\documentclass[prd,aps,nofootinbib,showpacs,preprintnumbers,amsmath,amssymb,floatfix]{revtex4}

\usepackage{times}
\usepackage{hyperref}

\usepackage{graphicx}% Include figure files
\usepackage{dcolumn}% Align table columns on decimal point
\usepackage{bm}% bold math
\usepackage{slashed}
\begin{document}

\title{Baryon quadrupole moment in the $1/N_c$ expansion of QCD}

\author{
V{\'\i}ctor Miguel Banda Guzm\'an
}
\affiliation{
Instituto de F{\'i}sica y Matem\'aticas, Universidad Michoacana de San Nicol\'as de Hidalgo, Edificio C-3, Apdo.\ Postal 2-82, Morelia, Michoac\'an, 58040, M\'exico
}

\author{
Rub\'en Flores-Mendieta
}
\affiliation{
Instituto de F{\'\i}sica, Universidad Aut\'onoma de San Luis Potos{\'\i}, \'Alvaro Obreg\'on 64, Zona Centro, San Luis Potos{\'\i}, S.L.P.\ 78000, Mexico
}

\author{
Johann Hern\'andez
}
\affiliation{
Instituto de F{\'\i}sica, Universidad Aut\'onoma de San Luis Potos{\'\i}, \'Alvaro Obreg\'on 64, Zona Centro, San Luis Potos{\'\i}, S.L.P.\ 78000, Mexico
}

\author{
Felipe de Jes\'us Rosales-Aldape
}
\affiliation{
Instituto de F{\'\i}sica, Universidad Aut\'onoma de San Luis Potos{\'\i}, \'Alvaro Obreg\'on 64, Zona Centro, San Luis Potos{\'\i}, S.L.P.\ 78000, Mexico
}

\date{\today}

\begin{abstract}
The quadrupole moments of ground state baryons are discussed in the framework of the $1/N_c$ expansion of QCD, where $N_c$ is the number of color charges. Theoretical expressions are first provided assuming an exact $SU(3)$ flavor symmetry, and then the effects of symmetry breaking are accounted for to linear order. The rather scarce experimental information available does not allow a detailed comparison 
between theory and experiment, so the free parameters in the approach are not determined. Instead, some useful new relations among quadrupole moments, valid even in the presence of first-order symmetry breaking, are provided. The overall predictions of the $1/N_c$ expansion are quite enlightening.
\end{abstract}

\pacs{12.39.Fe,11.15.Pg,13.40.Em,12.38.Bx}

\maketitle

\section{Introduction}

Understanding the structure of baryons is still a daunting task in quantum chromodynamics (QCD). The most interesting static properties of baryons, e.g., masses, magnetic moments, matter and charge radii, etc., fall in the nonperturbative regime of QCD so analytic calculations of these properties are not possible because the theory is strongly coupled at low energies, with no small expansion parameter.

The study of the electromagnetic properties of baryons is an active research area of both the theoretical and experimental bent. On the one hand, the analysis of the magnetic moments of baryons presents an opportunity to shed light on an accurate test of QCD and there are an important number of works on the subject; the approaches include, among others, the quark model (and its variants) \cite{qm1,qm2,qm3,qm4,qm5,qm6,qm7}, QCD sum rules \cite{sum1,sum2,sum4,sum5}, the $1/N_c$ expansion, where $N_c$ is the number of color charges \cite{jm94,dai,ji,lb,jenkins11}, chiral perturbation theory \cite{caldi,gasser,krause,jen92,milana,meiss,loyal,puglia,tib,geng,geng2}, the combined expansion in $1/N_c$ and chiral corrections \cite{rfm1,rfm2}, and lattice QCD \cite{latt1}, to name but a few. The experimental information available is robust \cite{part}, which allows detailed comparisons between theory and experiment.

On the other hand, the information about the higher-order electromagnetic moments (electric quadrupole and magnetic octupole moments) is less profuse. Analyses about the quadrupole moments of baryons have also been performed within the quark model (and its variants) \cite{kiv,sharma1,sharma2}, light cone QCD sum rules \cite{aliev,azizi}, the Skyrme model \cite{oh}, a QCD parametrization method \cite{bh1,bh2}, the $1/N_c$ expansion \cite{ji,bl1,bl2}, chiral perturbation theory \cite{butler,geng2}, and lattice QCD \cite{ca1,ca2}.
These lattice calculations are confined to the evaluation of the electromagnetic nucleon to $\Delta$ transition form factors. In contrast, the experimental data about quadrupole moments are rather scarce. The only experimental values reported are the helicity amplitudes for the process $\Delta^+ \to p \gamma$ \cite{part}, which can be used to extract the value of the ratio between the electric quadrupole ($E2$) and the magnetic moment ($M1$), $E2/M1$.

The present paper is focused on the computation of quadrupole moments of the ground state baryons in the context of the $1/N_c$ expansion,
which has proven to be quite effective in the calculation of static properties of baryons. Concrete examples can be found in the predictions for baryon masses \cite{djm95,lebed}, axial-vector \cite{djm95,dai,rfm98} and vector couplings \cite{rfm98,rfm04}. In all these scenarios, the approach gives a good description of the spin-flavor structure of QCD baryons with $N_c= 3$. For the purpose of the present paper, in Sec.~\ref{sec:sur} a survey of the $1/N_c$ expansion is presented to set the notation and conventions. In Sec.~\ref{sec:su3} the analysis starts with the construction of the most general spin-2, flavor octet operator which describes the baryon quadrupole moment. Next, the effects of flavor $SU(3)$ symmetry breaking (SB) to linear order are accounted for in Sec.~\ref{sec:sb}; the detailed construction of baryon operators which make up the series is described for each flavor representation present in the tensor product of the quadrupole moment and the perturbation to identify redundant operators. Once this task is completed, the full series is provided in Sec.~\ref{sec:full}. The lack of experimental information does not allow us to determine the free parameters of the theory so no attempt to predict any of the quadrupole moments numerically is made. Instead, various relations among them are provided. Some of them are valid in the limit of exact flavor symmetry, and others are valid even in the presence of SB. Some closing remarks are provided in Sec.~\ref{sec:clo}. The paper is complemented by the \hyperref[app:reduc]{Appendix~\ref*{app:reduc}}, where some useful baryon operator reductions are listed in order to discard redundant operators.

There are two papers that also address the evaluation of quadrupole moments of baryons within the $1/N_c$ expansion \cite{bl1,bl2}. In these papers, the minimal assumption of the single photon exchange ansatz is used, which implies that the photon probing these observables couples to only one quark line inside the baryon. This reduces the number of operators involved. This assumption is not used in the present analysis; instead the full operator basis is used here. Although this might be counterproductive due to the larger number of free parameters introduced, in fact, it leads to interesting relations among quadrupole moments which can not be determined otherwise. Another noticeable difference between the approach implemented in  Ref.~\cite{bl2} and the one used here concerns the way SB enters into play. In this reference SB is accounted for by modifying the spin-spin terms with ratios between the constituent quark masses. In the present analysis, as it was pointed out above, SB enters perturbatively. More details on the subject are provided in the following sections.

\section{\label{sec:sur}A survey of the $1/N_c$ expansion of QCD}

The present analysis builds on the seminal work on large-$N_c$ baryons presented in Ref.~\cite{djm95}, so only a few prominent facts will be highlighted here.

In the large-$N_c$ limit, the baryon sector exhibits a contracted $SU(2N_f)$ spin-flavor symmetry, where $N_f$ is the number of light quark flavors \cite{dm315,gs}. This baryon representation decomposes under $SU(2)\times SU(N_f)$ into a tower of baryon states with spins $J=1/2,3/2,\ldots,N_c/2$. The present analysis is done for the special case $N_f=3$. Therefore, the ground state baryons transform as the completely symmetric product of three $\mathbf{6}$'s of $SU(6)$, which is the $\mathbf{56}$ dimensional representation. Three spin $1/2$'s added together can yield spin $1/2$ or spin $3/2$, so the $\mathbf{56}$ representation contains spin-$1/2$ and spin-$3/2$ baryons.

Corrections to the large-$N_c$ limit are expressed in terms of $1/N_c$-suppressed operators with well-defined spin-flavor transformation properties \cite{gs}; this approach leads to the $1/N_c$ expansion of QCD.

Concretely, the $1/N_c$ expansion of a QCD operator at leading order reads \cite{djm95}
\begin{equation}
\mathcal{O}_\mathrm{QCD} = \sum_n c_n \frac{1}{N_c^{n-1}} \mathcal{O}_n,
\end{equation}
where the sum is over all possible operators $\mathcal{O}_n$, $n=0,\ldots,N_c$, which are polynomials in the spin-flavor generators of total order $n$. $\mathcal{O}_n$ are thus referred to as $n$-body operators. The spin-flavor generators can be written as 1-body quark operators acting on the $N_c$-quark baryon states, namely,
\begin{subequations}
\label{eq:su6gen}
\begin{eqnarray}
J^k & = & \sum_\alpha^{N_c} q_\alpha^\dagger \left(\frac{\sigma^k}{2}\otimes \openone \right) q_\alpha, \\
T^c & = & \sum_\alpha^{N_c} q_\alpha^\dagger \left(\openone \otimes \frac{\lambda^c}{2} \right) q_\alpha, \\
G^{kc} & = & \sum_\alpha^{N_c} q_\alpha^\dagger \left(\frac{\sigma^k}{2}\otimes \frac{\lambda^c}{2} \right) q_\alpha,
\end{eqnarray}
\end{subequations}
where $J^k$ are the spin generators, $T^c$ are the flavor generators, and $G^{kc}$ are the spin-flavor generators. The $SU(2N_f)$ spin-flavor generators satisfy the commutation relations listed in Table \ref{tab:surel} \cite{djm95}.
\begingroup
\begin{table}
\bigskip
\label{tab:su2fcomm}
\centerline{\vbox{ \tabskip=0pt \offinterlineskip
\halign{
\strut\quad $ # $\quad\hfil&\strut\quad $ # $\quad \hfil\cr
\multispan2\hfil $\left[J^i,T^a\right]=0,$ \hfil \cr
\noalign{\medskip}
\left[J^i,J^j\right]=i\epsilon^{ijk} J^k,
&\left[T^a,T^b\right]=i f^{abc} T^c,\cr
\noalign{\medskip}
\left[J^i,G^{ja}\right]=i\epsilon^{ijk} G^{ka},
&\left[T^a,G^{ib}\right]=i f^{abc} G^{ic},\cr
\noalign{\medskip}
\multispan2\hfil$\displaystyle [G^{ia},G^{jb}] = \frac{i}{4}\delta^{ij}
f^{abc} T^c + \frac{i}{2N_f} \delta^{ab} \epsilon^{ijk} J^k + \frac{i}{2} \epsilon^{ijk} d^{abc} G^{kc}.$ \hfill\cr
}}}
\caption{\label{tab:surel}$\mathrm{SU}(2 N_f)$ commutation relations.}
\end{table}
\endgroup
The operator coefficients $c_{n}$ also have power series expansions in $1/N_c$ beginning at order unity. On the other hand, $q_\alpha^\dagger$ and $q_\alpha$ are operators that create and annihilate states in the fundamental representation of $SU(6)$ and $\sigma^k$ and $\lambda^c$ are the Pauli spin and Gell-Mann flavor matrices, respectively. Because the baryon matrix elements of the spin-flavor generators (\ref{eq:su6gen}) can be taken as the values in the nonrelativistic quark model, this convention is usually referred to as the quark representation \cite{djm95}.

\section{\label{sec:su3}The quadrupole moment operator in the limit of exact $SU(3)$ flavor symmetry}

The electromagnetic current operator can be expanded in a power series of the photon momentum $k_\gamma$ (the multipole expansion); the series can be expressed as \cite{ji}
\begin{equation}
(J_\mathrm{em})^{ia} \propto \mu^{ia} + \mathcal{Q}^{(ij)a} k_\gamma^j + \ldots \label{eq:multipole}
\end{equation}
where $\mu^{ia}$ is the magnetic moment operator \cite{jm94} and $\mathcal{Q}^{(ij)a}$ is the quadrupole moment operator. $\mathcal{Q}^{(ij)a}$ is a spin-2 object and a flavor octet, so it transforms as $(2,\mathbf{8})$ under $SU(2)\times SU(3)$. $\mathcal{Q}^{(ij)a}$ is a symmetric, traceless tensor in the spin indices $i$ and $j$. The electromagnetic current is T-odd, so that $\mathcal{Q}^{(ij)a}$ is T-even.

The physical interpretation of the operators in the multipole expansion (\ref{eq:multipole}) can readily be seen through their matrix elements between $SU(6)$ symmetric baryon states. For example, the magnetic moment for a spin-$1/2$ baryon $B_p$, for a spin-$3/2$ baryon $B_q^\prime$, or for transitions $B_p \to B_q$ and $B_p^\prime \to B_q$, can be generically denoted by
\begin{equation}
\mu_{B} = \langle B|\mu^{3Q}|B \rangle, \label{eq:mmon}
\end{equation}
where the spin and flavor indices, $i$ and $a$, have been set to $3$ and $Q\equiv 3+(1/\sqrt{3})8$, respectively. Hereafter, any operator of the form $X^Q$ should be understood as $X^3 + (1/\sqrt{3})X^8$, where $X^3$ and $X^8$ denote the isovector and isoscalar components of the operator $X^a$, respectively.

Similarly, the zero component of the rank-2 tensor (in spin space) $\mathcal{Q}^{(ij)a}$, or equivalently, the $\ell=2$, $m_\ell=0$ component for $i=j=3$ and $a=Q$, is usually referred to as the \textit{spectroscopic} quadrupole moment \cite{bh1} for baryon $B_q^\prime$, which reads,
\begin{equation}
\mathcal{Q}_{B_q^\prime} = \langle B_q^\prime|\mathcal{Q}^{(33)Q}|B_q^\prime \rangle.
\end{equation}
Angular momentum selection rules forbid a spin-$1/2$ baryon from having a spectroscopic quadrupole moment. Similar definitions can also be given for transitions $B_p^\prime \to B_q$. For instance, for the radiative decay $\Delta^+ \to p\gamma$, the only multipoles that contribute are the magnetic moment ($M1$) and electric quadrupole ($E2$), which are defined as \cite{ji}
\begin{equation}
M1 = ek_\gamma^{1/2} \langle p|\mu^{3Q}|\Delta^+\rangle,
\end{equation}
and
\begin{equation}
E2 = \frac{1}{12} e k_\gamma^{3/2} \langle p|\mathcal{Q}^{(33)Q}|\Delta^+\rangle.
\end{equation}

On the other hand, to extract information about the shape of a spatially extended particle, its \textit{intrinsic} quadrupole moment $\mathcal{Q}_0$, given by
\begin{equation}
\mathcal{Q}_0 = \int d^3r \rho(r)(3z^2 - r^2),
\end{equation}
is generally used. $\mathcal{Q}_0$ is defined with respect to the body-fixed frame. For a charge density concentrated along the $z$-direction, $Q_0$ is positive and the particle is prolate. For a charge density concentrated in the equatorial plane perpendicular to $z$, $\mathcal{Q}_0$ is negative and the particle is oblate \cite{bh1}.

The intrinsic quadrupole moment must be distinguished from the spectroscopic one measured in the laboratory frame. Although a spin-1/2 baryon does not have a spectroscopic quadrupole moment, it may have an intrinsic one. Indeed, within various models, the proton and $\Delta^+$ are found to possess a prolate and an oblate deformation, respectively \cite{bh1}.

The present analysis is focused on the calculation of the spectroscopic quadrupole moment of baryons; it will be loosely referred to as the quadrupole moment hereafter. For this task, the quadrupole moment operator $\mathcal{Q}^{(ij)a}$ is constructed in the framework of the $1/N_c$ expansion, which requires to seek all the operator structures that transform as $(2,\mathbf{8})$ under $SU(2)\times SU(3)$, written as polynomials in the spin-flavor generators $J^i$, $T^a$, and $G^{ia}$ \cite{djm95}. Symmetry requirements are used in order to eliminate those structures which are not either T-even or symmetric or traceless in the spin indices. In this regard, operator structures should contain an even number of either $J$, $G$ or a combination of them, or an odd number of $J$, $G$ or a combination of them, along with a factor of $if^{abc}$ or $i\epsilon^{ijk}$ to yield Hermitian operators. Of course, spin and flavor indices must be properly saturated to have $(2,\mathbf{8})$ resultant operators. The $1$-body operator $i\epsilon^{ijk}G^{ka}$, for instance, is T-even and traceless but antisymmetric in $i$ and $j$, so it is discarded.

In the limit of exact $SU(3)$ flavor symmetry, the $1/N_c$ expansion of $\mathcal{Q}^{(ij)a}$ actually starts with the $2$-body operator,
\begin{equation}
O_2^{(ij)a} = \{J^i,G^{ja}\} + \{J^j,G^{ia}\} - \frac23 \delta^{ij} \{J^r,G^{ra}\}. \label{eq:2b8}
\end{equation}

At $3$-body operator level, several possibilities emerge, for instance,
\begin{equation}
\{T^a,\{J^i,J^j\}\}, \quad \{T^a,\{G^{ie},G^{je}\}\}, \quad \{G^{ia},\{T^e,G^{je}\}\}, \quad \{G^{ie},\{T^a,G^{je}\}\}, \quad d^{abc} \{J^i,\{T^b,G^{jc}\}\}.
\end{equation}

The use of operator identities \cite{djm95} restricts the number of linearly independent operators. In the \hyperref[app:reduc]{Appendix~\ref*{app:reduc}} some useful operator reductions are listed. Accordingly, a convenient $3$-body operator to include in the series is
\begin{equation}
O_3^{(ij)a} = \{T^a,\{J^i,J^j\}\} - \frac23 \delta^{ij} \{J^2,T^a\}, \label{eq:3b8}
\end{equation}
and all additional $3$-body operators are redundant and can be ignored.

On the other hand, $4$-body operators can be constructed as products of four $G$'s, two $J$'s and two $T$'s, two $J$'s and two $G$'s, three $G$'s and one $J$, and three $J$'s and one $G$, with the spin and flavor indices properly saturated. Among these structures, a convenient $4$-body operator is
\begin{equation}
O_4^{(ij)a} = \{\{J^i,J^j\},\{J^r,G^{ra}\}\} - \frac23 \delta^{ij} \{J^2,\{J^r,G^{ra}\}\}, \label{eq:4b8}
\end{equation}
whereas the others, according to the operator reductions listed in the \hyperref[app:reduc]{Appendix~\ref*{app:reduc}}, can be expressed in terms of either (\ref{eq:4b8}) and the lower-order operators (\ref{eq:2b8}) and (\ref{eq:3b8}), or in terms of (\ref{eq:2b8}) itself and its anticommutator with $J^2$. Hence, there is a second $4$-body operator obtained as
\begin{equation}
\tilde O_4^{(ij)a} = \{J^2,O_2^{(ij)a}\}.
\end{equation}

Now, $5$-body operators can be constructed out of the tensor products of the $4$-body operators listed above and $T^a$, with the proper contraction of spin and flavor indices. Following the operator reductions of the \hyperref[app:reduc]{Appendix~\ref*{app:reduc}}, it can be concluded that there is a single $5$-body operator given by
\begin{equation}
O_5^{(ij)a} = \{J^2,O_3^{(ij)a}\}.
\end{equation}

Therefore, without loss of generality, starting from $O_m^{(ij)a}$ ($m=2,3,4$), higher-order operators can be obtained as anticommutators of these operators with $J^2$, i.e., $O_{m+2}^{(ij)a}=\{J^2,O_m^{(ij)a}\}$, for $m\geq 3$. There are additional higher-order operators computed as $\tilde O_{n+2}^{(ij)a}=\{J^2,\tilde O_n^{(ij)a}\}$, for $n$ even, $n\geq 4$.

Thus, in the limit of exact $SU(3)$ flavor symmetry, the $1/N_c$ expansion of $\mathcal{Q}^{(ij)a}$ is written as
\begin{equation}
\mathcal{Q}^{(ij)a} = \sum_{m=2,3}^{N_c} \frac{1}{N_c^{m-1}} k_m O_m^{(ij)a} + \sum_{n=4,6}^{N_c-1} \frac{1}{N_c^{n-1}} \tilde k_n \tilde O_n^{(ij)a}, \label{eq:qsym}
\end{equation}
where $k_m, \tilde k_n$ are unknown parameters which also have a $1/N_c$ expansion beginning at order unity; these parameters are multiplied by a characteristic hadronic quadrupole size (in $\mathrm{fm}^2$). The vanishing trace condition can be easily verified as
\begin{equation}
\delta^{ij} \mathcal{Q}^{(ij)a} = \mathcal{Q}^{(ii)a} = 0.
\end{equation}

Expansion (\ref{eq:qsym}) can be truncated for arbitrary flavor $a$ after the first three operators $O_2^{(ij)a}$, $O_3^{(ij)a}$, and $O_4^{(ij)a}$ up to a relative correction of order $1/N_c^2$. For $N_c= 3$, only the two operators $O_2^{(ij)a}$ and $O_3^{(ij)a}$ are kept. Truncation beyond $O_3^{(ij)a}$ is justified for arbitrary $N_c$ up to a relative correction of order $1/N_c^2$ only when the physical baryons are under consideration. Thus, for $N_c=3$, the series reads
\begin{equation}
\mathcal{Q}^{(ij)a} = \frac{k_2}{N_c} \left[ \{J^i,G^{ja}\} + \{J^j,G^{ia}\} - \frac23 \delta^{ij}\{J^r,G^{ra}\} \right] + \frac{k_3}{N_c^2} \left[ \{T^a,\{J^i,J^j\}\} - \frac23 \delta^{ij} \{J^2,T^a\} \right], \label{eq:1onq}
\end{equation}
up to corrections of order $1/N_c^3$.

The quadrupole moments of baryons in the limit of exact $SU(3)$ flavor symmetry, $\mathcal{Q}_B^{SU(3)}$, can be obtained from the matrix elements of the baryon operators that make up $\mathcal{Q}^{(ij)a}$, Eq.~(\ref{eq:1onq}). These matrix elements for $N_c=3$ are listed in Tables \ref{t:aa1}, \ref{t:bb1}, and \ref{t:cc1}, for octet baryons, decuplet baryons, and decuplet-octet transitions. Although these matrix elements are provided for the special case $N_c=3$, its overall dependence on $N_c$ can be better seen from the matrix elements of the $1$-body operators $T^a$ and $G^{ia}$, $a=3$ and $a=8$, which occur quite often in the analysis. They can be rewritten in terms of the quark number and spin operators \cite{djm95},
\begin{subequations}
\label{eq:tgs}
\begin{equation}
T^8 = \frac{1}{2\sqrt{3}} (N_c-3N_s),
\end{equation}
\begin{equation}
G^{i8} = \frac{1}{2\sqrt{3}} (J^i-3J_s^i),
\end{equation}
\begin{equation}
T^3 = \frac12 (N_u-N_d),
\end{equation}
\begin{equation}
G^{i3} = \frac12 (J_u^i-J_d^i),
\end{equation}
\end{subequations}
where $N_c=N_u+N_d+N_s$ and $J^i=J_u^i+J_d^i+J_s^i$.

The leading $N_c$ counting of the matrix elements of operators (\ref{eq:tgs}) is deduced as follows: $T^8$, $G^{33}$, and $\{J^i,G^{i3}\}$ are order $\mathcal{O}(N_c)$; $J^3$, $T^3$, $G^{38}$, and $\{J^i,G^{i8}\}$ are order $\mathcal{O}(N_c^0)$. By using these counting rules and recalling that the operator coefficients $k_i$ are order unity, the isovector leading term in the series (\ref{eq:1onq}) is order $\mathcal{O}(N_c^0)$ and the subleading one is order $\mathcal{O}(N_c^{-2})$. The isoscalar leading and subleading terms, on the contrary, are both order $\mathcal{O}(N_c^{-1})$. Higher-order operators $O_n^{(ij)a}$ are suppressed by a relative factor of $1/N_c^2$ with respect to $O_{n-2}^{(ij)a}$.

With all the partial results properly gathered, the explicit computation of quadrupole moments can be carried out. An immediate result is that, for any octet baryon $B_p$,
\begin{equation}
\mathcal{Q}_{B_p}^{SU(3)} = 0,
\end{equation}
which is the sum of two null quantities (isovector and isoscalar quadrupole moments) and not the cancellation of two equal in magnitude but opposite in sign quantities. This is a completely expected (and consistent) result.

For decuplet baryons the quadrupole moments read
\begin{equation}
\mathcal{Q}_{B_q^\prime}^{SU(3)} = \frac49 q_{B_q^\prime} (k_2 + k_3), \label{eq:qsu3dec}
\end{equation}
which are valid up to a correction of order $1/N_c^3$. Here $q_{B_q^\prime}$ is the electric charge of decuplet baryon $B_q^\prime$ given by
\begin{equation}
q_{B_q^\prime} \equiv \langle B_q^\prime|T^Q|B_q^\prime\rangle.
\end{equation}

A few comments are in order here. Equation (\ref{eq:qsu3dec}) has been purposely written in a way to exhibit that the effects of higher-body operators can readily be accounted for without altering the basic structure of the equation itself. Thus one would be prompted to express Eq.~(\ref{eq:qsu3dec}) in terms of a single parameter, let us say $k$, so that
\begin{equation}
\mathcal{Q}_{B_q^\prime}^{SU(3)} = \frac49 q_{B_q^\prime} k,
\end{equation}
which would be valid to \textit{all orders} in the $1/N_c$ expansion. This approach, however, is not entirely correct. The reasons can be better seen when computing the transition quadrupole moments, $\mathcal{Q}_{B_q^\prime B_p}^{SU(3)}$. In this case,
\begin{subequations}
\begin{equation}
\mathcal{Q}_{\Delta^+p}^{SU(3)} = \frac{2\sqrt{2}}{9} k_2,
\end{equation}
\begin{equation}
\mathcal{Q}_{\Delta^0n}^{SU(3)} = \frac{2\sqrt{2}}{9} k_2,
\end{equation}
\begin{equation}
\mathcal{Q}_{{\Sigma^*}^0\Lambda}^{SU(3)} = \frac13 \sqrt{\frac23} k_2,
\end{equation}
\begin{equation}
\mathcal{Q}_{{\Sigma^*}^0\Sigma^0}^{SU(3)} = \frac{\sqrt{2}}{9} k_2,
\end{equation}
\begin{equation}
\mathcal{Q}_{{\Sigma^*}^+\Sigma^+}^{SU(3)} = \frac{2\sqrt{2}}{9} k_2,
\end{equation}
\begin{equation}
\mathcal{Q}_{{\Sigma^*}^-\Sigma^-}^{SU(3)} = 0, \label{eq:sssm}
\end{equation}
\begin{equation}
\mathcal{Q}_{{\Xi^*}^0\Xi^0}^{SU(3)} = \frac{2\sqrt{2}}{9} k_2,
\end{equation}
and
\begin{equation}
\mathcal{Q}_{{\Xi^*}^-\Xi^-}^{SU(3)} = 0, \label{eq:xisxim}
\end{equation}
\end{subequations}
which are valid up to a correction of order $1/N_c^2$. Due to the fact that the operator coefficients $k_i$ participate differently in $\mathcal{Q}_{B_q^\prime}^{SU(3)}$ and $\mathcal{Q}_{B_q^\prime B_p}^{SU(3)}$, there is not a unique way of recombining these operator coefficients to reduce the number of free parameters. Therefore, retaining up to 3-body operators in the $1/N_c$ expansion of quadrupole moments implies the existence of two free independent parameters, $k_2$ and $k_3$.

In the $SU(3)$ limit, two relations become evident, namely.
\begin{equation}
\frac12 \mathcal{Q}_{\Delta^{++}}^{SU(3)} = \mathcal{Q}_{\Delta^+}^{SU(3)} = -\mathcal{Q}_{\Delta^-}^{SU(3)} = \mathcal{Q}_{{\Sigma^*}^+}^{SU(3)} = -\mathcal{Q}_{{\Sigma^*}^-}^{SU(3)} = -\mathcal{Q}_{{\Xi^*}^-}^{SU(3)}, \label{eq:qm1}
\end{equation}
and
\begin{equation}
\mathcal{Q}_{\Delta^+p}^{SU(3)} = \mathcal{Q}_{\Delta^0n}^{SU(3)} = \mathcal{Q}_{{\Sigma^*}^+\Sigma^+}^{SU(3)} = \mathcal{Q}_{{\Xi^*}^0\Xi^0}^{SU(3)} = 2 \mathcal{Q}_{{\Sigma^*}^0\Sigma^0}^{SU(3)} = \frac{2}{\sqrt{3}} \mathcal{Q}_{{\Sigma^*}^0\Lambda}^{SU(3)}. \label{eq:qm2}
\end{equation}
The last relations have also been noticed within the chiral constituent quark model analysis of Ref.~\cite{sharma2}. As a side remark, notice that $\mathcal{Q}_{{\Sigma^*}^-\Sigma^-}^{SU(3)} = \mathcal{Q}_{{\Xi^*}^-\Xi^-}^{SU(3)} = 0$, which is a well-known result derived in the quark model \cite{sharma2}. These transitions are forbidden by $U$-spin conservation if flavor symmetry is exact.

Now, it is straightforward to test the combinations sensitive to $I=3$ and $I=2$ operators \cite{bl2}. For the former case,
\begin{subequations}
\begin{equation}
\mathcal{Q}_{\Delta^{++}}^{SU(3)} - 3 \mathcal{Q}_{\Delta^+}^{SU(3)} + 3 \mathcal{Q}_{\Delta^0}^{SU(3)} - \mathcal{Q}_{\Delta^-}^{SU(3)} = 0,
\end{equation}
and for the latter case,
\begin{equation}
\mathcal{Q}_{\Delta^{++}}^{SU(3)} - \mathcal{Q}_{\Delta^+}^{SU(3)} - \mathcal{Q}_{\Delta^0}^{SU(3)} + \mathcal{Q}_{\Delta^-}^{SU(3)} = 0,
\end{equation}
\begin{equation}
\mathcal{Q}_{{\Sigma^*}^+}^{SU(3)} - 2\mathcal{Q}_{{\Sigma^*}^0}^{SU(3)} + \mathcal{Q}_{{\Sigma^*}^-}^{SU(3)} = 0, \label{eq:isoS}
\end{equation}
\begin{equation}
\mathcal{Q}_{\Delta^+p}^{SU(3)} - \mathcal{Q}_{\Delta^0n}^{SU(3)} = 0, \label{eq:isoDN}
\end{equation}
\begin{equation}
\mathcal{Q}_{{\Sigma^*}^+\Sigma^+}^{SU(3)} - 2\mathcal{Q}_{{\Sigma^*}^0\Sigma^0}^{SU(3)} + \mathcal{Q}_{{\Sigma^*}^-\Sigma^-}^{SU(3)} = 0. \label{eq:isoST}
\end{equation}
\end{subequations}

\begin{table*}
\caption{\label{t:aa1}Matrix elements of baryon operators corresponding to quadrupole moments of octet baryons for $N_c=3$: $SU(3)$ case.}
\begin{ruledtabular}
\begin{tabular}{lcccccccc}
 &$n$ & $p$ & $\Sigma^-$ & $\Sigma^0$ & $\Sigma^+$ & $\Xi^-$ & $\Xi^0$ & $\Lambda$\\
\hline
$\langle \{J^3,G^{33}\} \rangle$ & $-\frac{5}{12}$ & $\frac{5}{12}$ & $-\frac13$ & $0$ & $\frac13$ & $\frac{1}{12}$ & $-\frac{1}{12}$ & $0$ \\
$\langle \delta^{33} \{J^r,G^{r3}\} \rangle$ & $-\frac54$ & $\frac54$ & $-1$ & $0$ & $1$ & $\frac14$ & $-\frac14$ & $0$ \\
$\langle \{T^3,\{J^3,J^3\}\} \rangle$ & $-\frac12$ & $\frac12$ & $-1$ & $0$ & $1$ & $-\frac12$ & $\frac12$ & $0$ \\
$\langle \delta^{33} \{J^2,T^3\} \rangle$ & $-\frac34$ & $\frac34$ & $-\frac32$ & $0$ & $\frac32$ & $-\frac34$ & $\frac34$ & $0$ \\
$\langle \{J^3,G^{38}\} \rangle$ & $\frac{1}{4\sqrt{3}}$ & $\frac{1}{4\sqrt{3}}$ & $\frac{1}{2\sqrt{3}}$ & $\frac{1}{2\sqrt{3}}$ & $\frac{1}{2 \sqrt{3}}$ & $-\frac{\sqrt{3}}{4}$ & $-\frac{\sqrt{3}}{4}$ & $-\frac{1}{2\sqrt{3}}$ \\
$\langle \delta^{33} \{J^r,G^{r8}\} \rangle$ & $\frac{\sqrt{3}}{4}$ & $\frac{\sqrt{3}}{4}$ & $\frac{\sqrt{3}}{2}$ & $\frac{\sqrt{3}}{2}$ & $\frac{\sqrt{3}}{2}$ & $-\frac{3\sqrt{3}}{4}$ & $-\frac{3\sqrt{3}}{4}$ & $-\frac{\sqrt{3}}{2}$ \\
$\langle \{T^8,\{J^3,J^3\}\} \rangle$ & $\frac{\sqrt{3}}{2}$ & $\frac{\sqrt{3}}{2}$ & $0$ & $0$ & $0$ & $-\frac{\sqrt{3}}{2}$ & $-\frac{\sqrt{3}}{2}$ & $0$ \\
$\langle \delta^{33}\{J^2,T^8\} \rangle$ & $\frac{3\sqrt{3}}{4}$ & $\frac{3\sqrt{3}}{4}$ & $0$ & $0$ & $0$ & $-\frac{3\sqrt{3}}{4}$ & $-\frac{3\sqrt{3}}{4}$ & $0$ \\
\end{tabular}
\end{ruledtabular}
\end{table*}

\begin{table*}
\caption{\label{t:bb1}Matrix elements of baryon operators corresponding to quadrupole moments of decuplet baryons for $N_c=3$: $SU(3)$ case.}
\begin{ruledtabular}
\begin{tabular}{lcccccccccc}
 & $\Delta^{++}$ & $\Delta^+$ & $\Delta^0$ & $\Delta^-$ & ${\Sigma^*}^+$ & ${\Sigma^*}^0$ & ${\Sigma^*}^-$ & ${\Xi^*}^-$ & ${\Xi^*}^0$ & $\Omega^-$ \\
\hline
$\langle \{J^3,G^{33}\} \rangle$ & $\frac94$ & $\frac34$ & $-\frac34$ & $-\frac94$ & $\frac32$ & $0$ & $-\frac32$ & $-\frac34$ & $\frac34$ & $0$ \\
$\langle \delta^{33} \{J^r,G^{r3}\} \rangle$ & $\frac{15}{4}$ & $\frac54$ & $-\frac54$ & $-\frac{15}{4}$ & $\frac52$ & $0$ & $-\frac52$ & $-\frac54$ & $\frac54$ & $0$ \\
$\langle \{T^3,\{J^3,J^3\}\} \rangle$ & $\frac{27}{2}$ & $\frac92$ & $-\frac92$ & $-\frac{27}{2}$ & $9$ & $0$ & $-9$ & $-\frac92$ & $\frac92$ & $0$ \\
$\langle \delta^{33} \{J^2,T^3\} \rangle$ & $\frac{45}{4}$ & $\frac{15}{4}$ & $-\frac{15}{4}$ & $-\frac{45}{4}$ & $\frac{15}{2}$ & $0$ & $-\frac{15}{2}$ & $-\frac{15}{4}$ & $\frac{15}{4}$ & $0$ \\
$\langle \{J^3,G^{38}\} \rangle$ & $\frac{3\sqrt{3}}{4}$ & $\frac{3\sqrt{3}}{4}$ & $\frac{3\sqrt{3}}{4}$ & $\frac{3\sqrt{3}}{4}$ & $0$ & $0$ & $0$ & $-\frac{3\sqrt{3}}{4}$ & $-\frac{3\sqrt{3}}{4}$ & $-\frac{3\sqrt{3}}{2}$ \\
$\langle \delta^{33} \{J^r,G^{r8}\} \rangle$ & $\frac{5\sqrt{3}}{4}$ & $\frac{5\sqrt{3}}{4}$ & $\frac{5\sqrt{3}}{4}$ & $\frac{5\sqrt{3}}{4}$ & $0$ & $0$ & $0$ & $-\frac{5\sqrt{3}}{4}$ & $-\frac{5\sqrt{3}}{4}$ & $-\frac{5\sqrt{3}}{2}$ \\
$\langle \{T^8,\{J^3,J^3\}\} \rangle$ & $\frac{9\sqrt{3}}{2}$ & $\frac{9\sqrt{3}}{2}$ & $\frac{9\sqrt{3}}{2}$ & $\frac{9\sqrt{3}}{2}$ & $0$ & $0$ & $0$ & $-\frac{9\sqrt{3}}{2}$ & $-\frac{9\sqrt{3}}{2}$ & $-9\sqrt{3}$ \\
$\langle \delta^{33}\{J^2,T^8\} \rangle$ & $\frac{15\sqrt{3}}{4}$ & $\frac{15\sqrt{3}}{4}$ & $\frac{15\sqrt{3}}{4}$ & $\frac{15\sqrt{3}}{4}$ & $0$ & $0$ & $0$ & $-\frac{15\sqrt{3}}{4}$ & $-\frac{15\sqrt{3}}{4}$ & $-\frac{15\sqrt{3}}{2}$ \\
\end{tabular}
\end{ruledtabular}
\end{table*}

\begin{table*}
\caption{\label{t:cc1}Matrix elements of baryon operators corresponding to transition quadrupole moments for $N_c=3$: $SU(3)$ case.}
\begin{ruledtabular}
\begin{tabular}{lcccccccc}
 & $\Delta^+p$ & $\Delta^0n$ & ${\Sigma^*}^0\Lambda$ & ${\Sigma^*}^0\Sigma^0$ & ${\Sigma^*}^+\Sigma^+$ & ${\Sigma^*}^-\Sigma^-$ & ${\Xi^*}^0\Xi^0$ & ${\Xi^*}^-\Xi^-$ \\
\hline
$\langle \{J^3,G^{33}\} \rangle$ & $\frac{\sqrt{2}}{3}$ & $\frac{\sqrt{2}}{3}$ & $\frac{1}{\sqrt{6}}$ & $0$ & $\frac{1}{3\sqrt{2}}$ & $-\frac{1}{3\sqrt{2}}$ & $\frac{1}{3\sqrt{2}}$ & $-\frac{1}{3\sqrt{2}}$ \\
$\langle \delta^{33} \{J^r,G^{r3}\} \rangle$ & $0$ & $0$ & $0$ & $0$ & $0$ & $0$ & $0$ & $0$ \\
$\langle \{T^3,\{J^3,J^3\}\} \rangle$ & $0$ & $0$ & $0$ & $0$ & $0$ & $0$ & $0$ & $0$ \\
$\langle \delta^{33} \{J^2,T^3\} \rangle$ & $0$ & $0$ & $0$ & $0$ & $0$ & $0$ & $0$ & $0$ \\
$\langle \{J^3,G^{38}\} \rangle$ &$0$ & $0$ & $0$ & $\frac{1}{\sqrt{6}}$ & $\frac{1}{\sqrt{6}}$ & $\frac{1}{\sqrt{6}}$ & $\frac{1}{\sqrt{6}}$ & $\frac{1}{\sqrt{6}}$ \\
$\langle \delta^{33} \{J^r,G^{r8}\} \rangle$ & $0$ & $0$ & $0$ & $0$ & $0$ & $0$ & $0$ & $0$ \\
$\langle \{T^8,\{J^3,J^3\}\} \rangle$ & $0$ & $0$ & $0$ & $0$ & $0$ & $0$ & $0$ & $0$ \\
$\langle \delta^{33}\{J^2,T^8\} \rangle$ & $0$ & $0$ & $0$ & $0$ & $0$ & $0$ & $0$ & $0$ \\
\end{tabular}
\end{ruledtabular}
\end{table*}

Additional expressions valid in the limit of exact $SU(3)$ symmetry can be extracted from Ref.~\cite{bh2}. Apart from the isospin relations (\ref{eq:isoS}), (\ref{eq:isoDN}), and (\ref{eq:isoST}), a and a vanishing $\mathcal{Q}_{\Delta^0}^{SU(3)}$, the expressions are given by
\begin{subequations}
\label{eq:bhrel}
\begin{equation}
\mathcal{Q}_{\Delta^-}^{SU(3)} + \mathcal{Q}_{\Delta^+}^{SU(3)} = 0,
\end{equation}
\begin{equation}
2 \mathcal{Q}_{\Delta^-}^{SU(3)} + \mathcal{Q}_{\Delta^{++}}^{SU(3)} = 0,
\end{equation}
\begin{equation}
3(\mathcal{Q}_{{\Xi^*}^-}^{SU(3)} - \mathcal{Q}_{{\Sigma^*}^-}^{SU(3)}) - (\mathcal{Q}_{\Omega^-}^{SU(3)} - \mathcal{Q}_{\Delta^-}^{SU(3)}) = 0,
\end{equation}
\begin{equation}
\mathcal{Q}_{\Delta^-}^{SU(3)} - \mathcal{Q}_{{\Sigma^*}^-}^{SU(3)} - \sqrt{2} \mathcal{Q}_{{\Sigma^*}^- \Sigma^-}^{SU(3)} = 0,
\end{equation}
\begin{equation}
\mathcal{Q}_{\Delta^+}^{SU(3)} - \mathcal{Q}_{{\Sigma^*}^+}^{SU(3)} + \sqrt{2} \mathcal{Q}_{\Delta^+ p}^{SU(3)} - \sqrt{2} \mathcal{Q}_{{\Sigma^*}^+ \Sigma^+}^{SU(3)} = 0,
\end{equation}
\begin{equation}
\mathcal{Q}_{{\Sigma^*}^0}^{SU(3)} - \frac{1}{\sqrt{2}} \mathcal{Q}_{{\Sigma^*}^0 \Sigma^0}^{SU(3)} + \frac{1}{\sqrt{6}} \mathcal{Q}_{{\Sigma^*}^0 \Lambda}^{SU(3)} = 0,
\end{equation}
\begin{equation}
\mathcal{Q}_{{\Sigma^*}^-}^{SU(3)} - \mathcal{Q}_{{\Xi^*}^-}^{SU(3)} - \frac{1}{\sqrt{2}} \mathcal{Q}_{{\Xi^*}^- \Xi^-}^{SU(3)} - \frac{1}{\sqrt{2}} \mathcal{Q}_{{\Sigma^*}^- \Sigma^-}^{SU(3)} = 0,
\end{equation}
\begin{equation}
\mathcal{Q}_{{\Xi^*}^0}^{SU(3)} + \frac{1}{\sqrt{2}} \mathcal{Q}_{{\Xi^*}^0 \Xi^0}^{SU(3)} - \sqrt{\frac23} \mathcal{Q}_{{\Sigma^*}^0 \Lambda}^{SU(3)} = 0,
\end{equation}
\end{subequations}
which are also verified with the expressions obtained within the formalism presented here. Relations (\ref{eq:bhrel}a) and (\ref{eq:bhrel}b) are easily explained as a consequence of $\mathcal{Q}_{B_q^\prime}$ being proportional to the electric charge of baryon $B_q^\prime$. 

\section{\label{sec:sb}$\mathcal{Q}^{(ij)a}$ with first-order $SU(3)$ symmetry breaking}

In the Standard Model, flavor $SU(3)$ symmetry breaking is given by the current quark mass term in the QCD Lagrangian with $m_u, m_d \ll m_s$ and transforms as a flavor octet.

The correction to $\mathcal{Q}^{(ij)a}$ is obtained to linear order in SB from the tensor product of the quadrupole moment and the perturbation, which transform under $SU(2)\times SU(3)$ as $(2,\mathbf{8})$ and $(0,\mathbf{8})$, respectively. The representations contained in this tensor product are $(2,\mathbf{1})$, $(2,\mathbf{8})$, $(2,\mathbf{8})$, $(2,\mathbf{10}+\overline{\mathbf{10}})$, and $(2,\mathbf{27})$. The task of finding $1/N_c$ operator expansions for these representations is dealt with in the following subsections.

\subsection{$(2,\mathbf{1})$}

There is only a 0-body operator transforming as $(2,\mathbf{1})$ under $SU(2)\times SU(3)$,
\begin{equation}
O_0^{ij} = \delta^{ij}\openone,
\end{equation}
and a single $1$-body operator
\begin{equation}
O_1^{ij} = i\epsilon^{ijk}J^k,
\end{equation}
none of which contributes to $\mathcal{Q}^{(ij)a}$ by virtue of the symmetry or trace conditions discussed above. Thus, the only nontrivial operator found is the $2$-body operator,
\begin{equation}
O_2^{ij} = \{J^i,J^j\} - \frac23 \delta^{ij} J^2,
\end{equation}
because even the $3$-body operator,
\begin{equation}
\{J^i,\{T^e,G^{je}\}\},
\end{equation}
is also redundant, according to Eq.~(\ref{eq:jtg}) of the \hyperref[app:reduc]{Appendix~\ref*{app:reduc}}.

Therefore, the SB contribution to $\mathcal{Q}^{(ij)a}$ from the $(2,\mathbf{1})$ representation reads\footnote{Hereafter, for the ease of notation, $O_{n,\mathbf{rep}}^{(ij)a}$ will stand for an $n$-body operator belonging to flavor representation $\mathbf{rep}$.}
\begin{equation}
O_{2,\mathbf{1}}^{(ij)a} = \delta^{a8} \{J^i,J^j\} - \frac23 \delta^{ij} \delta^{a8} J^2, \label{eq:212}
\end{equation}
whereas higher-order operators are consecutively obtained as $O_{2m+2,\mathbf{1}}^{(ij)a} = \{J^2,O_{2m,\mathbf{1}}^{(ij)a}\}$ for $m\geq 1$.

\subsection{$(2,\mathbf{8})$}

The $(2,\mathbf{8})$ operators that generate SB corrections to $\mathcal{Q}^{(ij)a}$ can be obtained in a close analogy with expressions (\ref{eq:2b8}) and (\ref{eq:3b8}) and read respectively for $2$- and $3$-body operators,
\begin{equation}
O_{2,\mathbf{8}}^{(ij)a} = d^{ae8} (\{J^i,G^{je}\} + \{J^j,G^{ie}\}) - \frac23 \delta^{ij} d^{ae8} \{J^r,G^{re}\}, \label{eq:282}
\end{equation}
and
\begin{equation}
O_{3,\mathbf{8}}^{(ij)a} = d^{ae8} \{T^e,\{J^i,J^j\}\} - \frac23 \delta^{ij} d^{ae8} \{J^2,T^e\}. \label{eq:283}
\end{equation}
Higher-order operators are obtained as $O_{n+2,\mathbf{8}}^{(ij)a} = \{J^2,O_{n,\mathbf{8}}^{(ij)a}\}$ for $n \geq 2$.

Additional $(2,\mathbf{8})$ operators obtained by replacing the $d^{ab8}$ symbol with the $if^{ab8}$ symbol turn out to be T-odd so they are forbidden by time reversal invariance.

\subsection{$(2,\mathbf{27})$}

The analysis of $(2,\mathbf{27})$ operators is more involved than the ones previously discussed. In the present case, not only the symmetry of operators under the exchange of spin indices must be manifest, but also the symmetry of operators under the exchange of flavor indices, which is required for flavor-$\mathbf{27}$ operators. Let $S_+^{(ij)\{ab\}}$ be one of such operators. The singlet and octet components are subtracted off $S_+^{(ij)\{ab\}}$ to obtain a genuine flavor-$\mathbf{27}$ operator $S^{(ij)\{ab\}}$ according to \cite{djm95}
\begin{equation}
S^{(ij)\{ab\}} = S_+^{(ij)\{ab\}} - \frac{1}{N_f^2-1} \delta^{ab} S_+^{(ij)\{ee\}} - \frac{N_f}{N_f^2-4} d^{abe} d^{ghe}S_+^{(ij)\{gh\}}. \label{eq:227form}
\end{equation}
The contribution of $S^{(ij)\{ab\}}$ to $\mathcal{Q}^{(ij)a}$ is actually obtained by setting $b=8$ so the contribution is effectively $S^{(ij)\{a8\}}$.

The $1/N_c$ expansion for a $(2,\mathbf{27})$ operator that contributes to $\mathcal{Q}^{(ij)a}$ begins with a single $2$-body operator
\begin{equation}
\{G^{ia},G^{jb}\} + \{G^{ja},G^{ib}\} - \frac23 \delta^{ij} \{G^{ra},G^{rb}\},
\end{equation}
which, after subtracting singlet and octet components in accordance with prescription (\ref{eq:227form}), gets the form
\begin{eqnarray}
&  & \{G^{ia},G^{jb}\} + \{G^{ja},G^{ib}\} - \frac{2}{N_f^2-1} \delta^{ab} \{G^{ie},G^{je}\} - \frac{2N_f}{N_f^2-4} d^{abe}d^{egh} \{G^{ig},G^{jh}\} - \frac23 \delta^{ij} \{G^{ra},G^{rb}\} \nonumber \\
&  & \mbox{} + \frac23 \frac{1}{N_f^2-1} \delta^{ij} \delta^{ab} \{G^{re},G^{re}\} + \frac23 \frac{N_f}{N_f^2-4} \delta^{ij} d^{abe}d^{egh} \{G^{rg},G^{rh}\}.
\end{eqnarray}

After a straightforward algebraic manipulation, the SB correction to $\mathcal{Q}^{(ij)a}$ from this $2$-body operator becomes
\begin{eqnarray}
O_{2,\mathbf{27}}^{(ij)a} & = & \{G^{ia},G^{j8}\} + \{G^{i8},G^{ja}\} + \frac16 \delta^{ij} \{T^{a},T^{8}\} + \frac13 \delta^{ij}f^{age} f^{8he}\{T^g,T^h\} - \frac{1}{N_f+2} d^{ae8} (\{J^{i},G^{je}\} + \{J^{j},G^{ie}\}) \nonumber \\
&  & \mbox{} - \frac13 \frac{N_f}{N_f+2} \delta^{ij} d^{ae8} \{J^{r},G^{re}\} - \frac{1}{N_f(N_f+1)} \delta^{a8} \{J^{i},J^{j}\} - \frac23 \frac{1}{N_f+1} \delta^{ij} \delta^{a8} J^2 - \frac16 (N_c+N_f) \delta^{ij} d^{ae8} T^{e} \nonumber \\
&  & \mbox{} - \frac{N_c(N_c+2N_f)}{6N_f} \delta^{ij} \delta^{a8}. \label{eq:2272}
\end{eqnarray}

As for $3$-body operators, there is a single one given by
\begin{equation}
\frac12 \{T^a,\{J^i,G^{jb}\} + \{J^j,G^{ib}\}\} + \frac12 \{T^b,\{J^i,G^{ja}\} + \{J^j,G^{ia}\}\} - \frac13 \delta^{ij} (\{T^a,\{J^r,G^{rb}\}\} + \{T^b,\{J^r,G^{ra}\}\}),
\end{equation}
which turns into
\begin{eqnarray}
&  & \frac12 \{T^{a},\{J^{i},G^{jb}\} + \{J^{j},G^{ib}\}\} + \frac12 \{T^{b},\{J^{i},G^{ja}\} + \{J^{j},G^{ia}\}\} - \frac{1}{N_f^2-1} \delta^{ab} \{T^{e},\{J^{i},G^{je}\} + \{J^{j},G^{ie}\}\} \nonumber \\
&  & \mbox{} - \frac{N_f}{N_f^2-4} d^{abe} d^{fge} \{T^{f},\{J^{i},G^{jg}\} + \{J^{j},G^{ig}\}\} - \frac13 \delta^{ij} \{T^{a},\{J^{r},G^{rb}\}\} - \frac13 \delta^{ij} \{T^{b},\{J^{r},G^{ra}\}\} \nonumber \\
&  & \mbox{} + \frac23 \frac{1}{N_f^2-1} \delta^{ij} \delta^{ab} \{T^{e},\{J^{r},G^{re}\}\} + \frac23 \frac{N_f}{N_f^2-4} \delta^{ij} d^{abe} d^{fge} \delta^{ij} \{T^{f},\{J^{r},G^{rg}\}\},
\end{eqnarray}
by following the prescription (\ref{eq:227form}). Therefore, the SB correction to $\mathcal{Q}^{(ij)a}$ from $3$-body operators is then given by
\begin{eqnarray}
O_{3,\mathbf{27}}^{(ij)a} & = & \frac12 \{T^{a},\{J^{i},G^{j8}\} + \{J^{j},G^{i8}\}\} + \frac12 \{T^{8},\{J^{i},G^{ja}\} + \{J^{j},G^{ia}\}\} - \frac13 \delta^{ij} \{T^{a},\{J^{r},G^{r8}\}\} - \frac13 \delta^{ij} \{T^{8},\{J^{r},G^{ra}\}\} \nonumber \\
&  & \mbox{} - \frac{1}{N_f+2} d^{ae8} \{T^{e},\{J^{i},J^{j}\}\} + \frac23 \frac{1}{N_f+2} \delta^{ij} d^{ae8} \{J^2,T^{e}\} - \frac{N_c+N_f}{N_f+2} d^{ae8} (\{J^{i},G^{je}\} + \{J^{j},G^{ie}\}) \nonumber \\
&  & \mbox{} + \frac23 \frac{N_c+N_f}{N_f+2} \delta^{ij} d^{ae8} \{J^{r},G^{re}\} - \frac{2(N_c+N_f)}{N_f(N_f+1)} \delta^{a8} \{J^{i},J^{j}\} + \frac43 \frac{N_c+N_f}{N_f(N_f+1)} \delta^{ij} \delta^{a8} J^2. \label{eq:2273}
\end{eqnarray}

Next $4$-body operators can be worked out. They can conveniently constructed as tensor products of a spin-0 and a spin-1, $2$-body operators. There are two of such operators, namely,
\begin{equation}
\{\{J^i,J^j\},\{G^{ra},G^{rb}\}\} - \frac23 \delta^{ij} \{J^2,\{G^{ra},G^{rb}\}\},
\end{equation}
and
\begin{equation}
\{\{J^i,J^j\},\{T^a,T^b\}\} - \frac23 \delta^{ij} \{J^2,\{T^a,T^b\}\}.
\end{equation}

The procedure to subtract singlet and octet components yields, for the former,
\begin{eqnarray}
&  & \{\{J^i,J^j\},\{G^{ra},G^{rb}\}\} - \frac{1}{N_f^2-1} \delta^{ab}\{\{J^i,J^j\},\{G^{re},G^{re}\}\} - \frac{N_f}{N_f^2-4} d^{abe} d^{ghe} \{\{J^i,J^j\},\{G^{rg},G^{rh}\}\} \nonumber \\
&  & \mbox{} - \frac23 \delta^{ij} \{J^2,\{G^{ra},G^{rb}\}\} + \frac23 \frac{N_f}{N_f^2-4} \delta^{ij} d^{abe} d^{ghe} \{J^2,\{G^{rg},G^{rh}\}\} + \frac23 \frac{1}{N_f^2-1} \delta^{ij} \delta^{ab} \{J^2,\{G^{re},G^{re}\}\},
\end{eqnarray}
and for the latter,
\begin{eqnarray}
&  & \{\{J^i,J^j\},\{T^a,T^b\}\} - \frac{1}{N_f^2-1} \delta^{ab} \{\{J^i,J^j\},\{T^e,T^e\}\} - \frac{N_f}{N_f^2-4} d^{abe} d^{ghe} \{\{J^i,J^j\},\{T^g,T^h\}\} - \frac23 \delta^{ij}\{J^2,\{T^a,T^b\}\} \nonumber \\
&  & \mbox{} + \frac23 \frac{1}{N_f^2-1} \delta^{ij} \delta^{ab} \{J^2,\{T^e,T^e\}\} + \frac23 \frac{N_f}{N_f^2-4} \delta^{ij} d^{abe} d^{ghe} \{J^2,\{T^g,T^h\}\}.
\end{eqnarray}

Finally, the SB contributions from $4$-body operators to $\mathcal{Q}^{(ij)a}$ are given by
\begin{eqnarray}
O_{4,\mathbf{27}}^{(ij)a} & = & \{\{J^i,J^j\},\{G^{ra},G^{r8}\}\} + \frac12 \frac{N_f+2}{N_f(N_f^2-1)} \delta^{a8} \{J^2,\{J^i,J^j\}\} + \frac12 \frac{N_f+4}{N_f^2-4} d^{ae8} \{\{J^i,J^j\},\{J^r,G^{re}\}\} \nonumber \\
&  & \mbox{} - \frac34 \frac{(N_c+N_f)N_f}{N_f^2-4} d^{ae8} \{T^e,\{J^i,J^j\}\} - \frac34 \frac{N_c(N_c+2 N_f)}{N_f^2-1} \delta^{a8}\{J^i,J^j\} - \frac23 \delta^{ij}\{J^2,\{G^{ra},G^{r8}\}\} \nonumber \\
&  & \mbox{} - \frac13 \frac{N_f+2}{N_f(N_f^2-1)} \delta^{ij} \delta^{a8} \{J^2,J^2\} - \frac13 \frac{N_f+4}{N_f^2-4} \delta^{ij} d^{ae8}\{J^2,\{J^r,G^{re}\}\} + \frac12 \frac{(N_c+N_f)N_f}{N_f^2-4} \delta^{ij} d^{ae8} \{J^2,T^e\} \nonumber \\
&  & \mbox{} + \frac12 \frac{N_c(N_c+2N_f)}{N_f^2-1} \delta^{ij}\delta^{a8} J^2,
\end{eqnarray}
and
\begin{eqnarray}
\tilde O_{4,\mathbf{27}}^{(ij)a} & = & \{\{J^i,J^j\},\{T^a,T^8\}\} - \frac{2}{N_f^2-1} \delta^{a8} \{J^2,\{J^i,J^j\}\} - \frac{2N_f}{N_f^2-4} d^{ae8} \{\{J^i,J^j\},\{J^r,G^{re}\}\} \nonumber \\
&  & \mbox{} - \frac{(N_c+N_f)(N_f-4)}{N_f^2-4}d^{ae8} \{T^e,\{J^i,J^j\}\} - \frac{N_c(N_c+2 N_f)(N_f-2)}{N_f(N_f^2-1)}\delta^{a8}\{J^i,J^j\} - \frac23 \delta^{ij}\{J^2,\{T^a,T^8\}\} \nonumber \\
&  & \mbox{} + \frac43 \frac{1}{N_f^2-1} \delta^{ij} \delta^{a8} \{J^2,J^2\} + \frac43 \frac{N_f}{N_f^2-4} \delta^{ij} d^{ae8} \{J^2,\{J^r,G^{re}\}\} + \frac23 \frac{(N_c+N_f)(N_f-4)}{N_f^2-4} \delta^{ij} d^{ae8} \{J^2,T^e\} \nonumber \\
&  & \mbox{} + \frac23 \frac{N_c(N_c+2N_f)(N_f-2)}{N_f(N_f^2-1)} \delta^{ij} \delta^{a8} J^2 .
\end{eqnarray}

There is an additional $4$-body operator constructed as
\begin{equation}
\frac12 \{\{J^r,J^r\},\{G^{ia},G^{jb}\} + \{G^{ja},G^{ib}\}\},
\end{equation}
but it can be rewritten in terms of $\{J^2,O_2^{(ij)a}\}$, so it is redundant and can be discarded.

To close this section, notice that the flavor singlet and octet components subtracted off the original flavor $\mathbf{27}$ operators could have been respectively merged into the already defined $O_{n,\mathbf{1}}^{(ij)a}$ and $O_{n,\mathbf{8}}^{(ij)a}$ operators. The reason to keep these components in the original expression is twofold. First, the vanishing trace condition is kept in the full expression, and second, this allows us to disentangle the effects of different representations, so the corresponding operator coefficients parametrize pure $\mathbf{27}$ effects.

\subsection{$(2,\mathbf{10}+\overline{\mathbf{10}})$}

Contrary to the previous case, $(2,\mathbf{10}+\overline{\mathbf{10}})$ operators must be antisymmetric under the exchange of flavor indices, retaining the symmetry under the exchange of spin indices. Let $A_-^{(ij)[ab]}$ be one of such operators. Thus, in order to get a genuine flavor $\mathbf{10}+\overline{\mathbf{10}}$ operator $A^{(ij)[ab]}$, the flavor octet component must be subtracted off according to \cite{djm95}
\begin{equation}
A^{(ij)[ab]} = A_-^{(ij)[ab]} - \frac{1}{N_f} f^{abe}f^{ghe} A_-^{(ij)[gh]},
\end{equation}
where, by construction,
\begin{equation}
f^{abc} A^{(ij)[ab]} = 0. \label{eq:c10}
\end{equation}

With the above considerations, the series for the $(2,\mathbf{10}+\overline{\mathbf{10}})$ SB operators actually begins with a single $3$-body operator,
\begin{equation}
\frac12 \{T^a,\{J^i,G^{jb}\} + \{J^j,G^{ib}\}\} - \frac12 \{T^b,\{J^i,G^{ja}\} + \{J^j,G^{ia}\}\} - \frac13 \delta^{ij} (\{T^a,\{J^r,G^{rb}\}\} - \{T^b,\{J^r,G^{ra}\}\}),
\end{equation}
where the octet component to be subtracted off reads
\begin{equation}
\frac{1}{N_f} if^{abe}[J^2,\{J^i,G^{je}\}+\{J^j,G^{ie}\}].
\end{equation}
This last term is particularly interesting. First, notice that it is T-odd, so in principle it is forbidden in the $1/N_c$ expansion of $\mathcal{Q}^{(ij)a}$. And secondly, it vanishes under contraction of spin indices. It is necessary, though, to fulfill condition (\ref{eq:c10}).

Thus, the SB contribution to $\mathcal{Q}^{(ij)a}$ from $(2,\mathbf{10}+\overline{\mathbf{10}})$ operators is
\begin{eqnarray}
O_{3,\mathbf{10}+\overline{\mathbf{10}}}^{(ij)a} & = & \frac12 \{T^a,\{J^i,G^{j8}\} + \{J^j,G^{i8}\}\} - \frac12 \{T^8,\{J^i,G^{ja}\} + \{J^j,G^{ia}\}\} - \frac13 \delta^{ij} \{T^a,\{J^r,G^{r8}\}\} \nonumber \\
&  & \mbox{} + \frac13 \delta^{ij} \{T^8,\{J^r,G^{ra}\}\}, \label{eq:2103}
\end{eqnarray}
where the octet component has been safely ignored. Higher-order operators are obtained as $O_{n+2,\mathbf{10}+\overline{\mathbf{10}}}^{(ij)a} = \{J^2,O_{n,\mathbf{10}+\overline{\mathbf{10}}}^{(ij)a}\}$, for $n\geq 3$.

\begin{table*}
\caption{\label{t:aa2}Matrix elements of baryon operators corresponding to quadrupole moments of octet baryons for $N_c=3$: broken $SU(3)$ case.}
\begin{ruledtabular}
\begin{tabular}{lcccccccc}
 & $n$ & $p$ & $\Sigma^-$ & $\Sigma^0$ & $\Sigma^+$ & $\Xi^-$ & $\Xi^0$ & $\Lambda$ \\
\hline
$\langle \delta^{38} \{J^3,J^3\} \rangle$ & $0$ & $0$ & $0$ & $0$ & $0$ & $0$ & $0$ & $0$ \\
$\langle \delta^{33} \delta^{38} J^2 \rangle$ & $0$ & $0$ & $0$ & $0$ & $0$ & $0$ & $0$ & $0$ \\
$\langle d^{3e8} \{J^3,G^{3e}\} \rangle$ & $-\frac{5}{12\sqrt{3}}$ & $\frac{5}{12\sqrt{3}}$ & $-\frac{1}{3\sqrt{3}}$ & $0$ & $\frac{1}{3\sqrt{3}}$ & $\frac{1}{12\sqrt{3}}$ & $-\frac{1}{12\sqrt{3}}$ & $0$ \\
$\langle \delta^{33} d^{3e8} \{J^r,G^{re}\} \rangle$ & $-\frac{5}{4\sqrt{3}}$ & $\frac{5}{4\sqrt{3}}$ & $-\frac{1}{\sqrt{3}}$ & $0$ & $\frac{1}{\sqrt{3}}$ & $\frac{1}{4\sqrt{3}}$ & $-\frac{1}{4\sqrt{3}}$ & $0$ \\
$\langle d^{3e8} \{T^e,\{J^3,J^3\}\} \rangle$ & $-\frac{1}{2\sqrt{3}}$ & $\frac{1}{2\sqrt{3}}$ & $-\frac{1}{\sqrt{3}}$ & $0$ & $\frac{1}{\sqrt{3}}$ & $-\frac{1}{2\sqrt{3}}$ & $\frac{1}{2\sqrt{3}}$ & $0$ \\
$\langle \delta^{33} d^{3e8} \{J^2,T^e\} \rangle$ & $-\frac{\sqrt{3}}{4}$ & $\frac{\sqrt{3}}{4}$ & $-\frac{\sqrt{3}}{2}$ & $0$ & $\frac{\sqrt{3}}{2}$ & $-\frac{\sqrt{3}}{4}$ & $\frac{\sqrt{3}}{4}$ & $0$ \\
$\langle \{G^{33},G^{38}\} \rangle$ & $-\frac{5}{24\sqrt{3}}$ & $\frac{5}{24\sqrt{3}}$ & $-\frac{2}{3\sqrt{3}}$ & $0$ & $\frac{2}{3\sqrt{3}}$ & $-\frac{11}{24\sqrt{3}}$ & $\frac{11}{24\sqrt{3}}$ & $0$ \\
$\langle \delta^{33} \{T^3,T^8\} \rangle$ & $-\frac{\sqrt{3}}{2}$ & $\frac{\sqrt{3}}{2}$ & $0$ & $0$ & $0$ & $\frac{\sqrt{3}}{2}$ & $-\frac{\sqrt{3}}{2}$ & $0$ \\
$\langle \delta^{33} f^{3ge} f^{8he}\{T^g,T^h\} \rangle$ & $-\frac{\sqrt{3}}{4}$ & $\frac{\sqrt{3}}{4}$ & $0$ & $0$ & $0$ & $\frac{\sqrt{3}}{4}$ & $-\frac{\sqrt{3}}{4}$ & $0$ \\
$\langle \delta^{33} d^{3e8} T^e \rangle$ & $-\frac{1}{2\sqrt{3}}$ & $\frac{1}{2\sqrt{3}}$ & $-\frac{1}{\sqrt{3}}$ & $0$ & $\frac{1}{\sqrt{3}}$ & $-\frac{1}{2\sqrt{3}}$ & $\frac{1}{2\sqrt{3}}$ & $0$ \\
$\langle \delta^{33} \delta^{38} \rangle$ & $0$ & $0$ & $0$ & $0$ & $0$ & $0$ & $0$ & $0$ \\
$\langle \{T^3,\{J^3,G^{38}\}\} \rangle$ & $-\frac{1}{4\sqrt{3}}$ & $\frac{1}{4\sqrt{3}}$ & $-\frac{1}{\sqrt{3}}$ & $0$ & $\frac{1}{\sqrt{3}}$ & $\frac{\sqrt{3}}{4}$ & $-\frac{\sqrt{3}}{4}$ & $0$ \\
$\langle \{T^8,\{J^3,G^{33}\}\} \rangle$ & $-\frac{5}{4\sqrt{3}}$ & $\frac{5}{4\sqrt{3}}$ & $0$ & $0$ & $0$ & $-\frac{1}{4\sqrt{3}}$ & $\frac{1}{4\sqrt{3}}$ & $0$ \\
$\langle \delta^{33} \{T^3,\{J^r,G^{r8}\}\} \rangle$ & $-\frac{\sqrt{3}}{4}$ & $\frac{\sqrt{3}}{4}$ & $-\sqrt{3}$ & $0$ & $\sqrt{3}$ & $\frac{3\sqrt{3}}{4}$ & $-\frac{3\sqrt{3}}{4}$ & $0$ \\
$\langle \delta^{33} \{T^8,\{J^r,G^{r3}\}\} \rangle$ & $-\frac{5\sqrt{3}}{4}$ & $\frac{5\sqrt{3}}{4}$ & $0$ & $0$ & $0$ & $-\frac{\sqrt{3}}{4}$ & $\frac{\sqrt{3}}{4}$ & $0$ \\
$\langle \delta^{88} \{J^3,J^3\} \rangle$ & $\frac12$ & $\frac12$ & $\frac12$ & $\frac12$ & $\frac12$ & $\frac12$ & $\frac12$ & $\frac12$ \\
$\langle \delta^{33} \delta^{88} J^2 \rangle$ & $\frac34$ & $\frac34$ & $\frac34$ & $\frac34$ & $\frac34$ & $\frac34$ & $\frac34$ & $\frac34$ \\
$\langle d^{8e8} \{J^3,G^{3e}\} \rangle$ & $-\frac{1}{12}$ & $-\frac{1}{12}$ & $-\frac16$ & $-\frac16$ & $-\frac16$ & $\frac14$ & $\frac14$ & $\frac16$ \\
$\langle \delta^{33} d^{8e8} \{J^r,G^{re}\} \rangle$ & $-\frac14$ & $-\frac14$ & $-\frac12$ & $-\frac12$ & $-\frac12$ & $\frac34$ & $\frac34$ & $\frac12$ \\
$\langle d^{8e8} \{T^e,\{J^3,J^3\}\} \rangle$ & $-\frac12$ & $-\frac12$ & $0$ & $0$ & $0$ & $\frac12$ & $\frac12$ & $0$ \\
$\langle \delta^{33} d^{8e8} \{J^2,T^e\} \rangle$ & $-\frac34$ & $-\frac34$ & $0$ & $0$ & $0$ & $\frac34$ & $\frac34$ & $0$ \\
$\langle \{G^{38},G^{38}\} \rangle$ & $\frac{1}{24}$ & $\frac{1}{24}$ & $\frac12$ & $\frac12$ & $\frac12$ & $\frac{17}{24}$ & $\frac{17}{24}$ & $\frac16$ \\
$\langle \delta^{33} \{T^8,T^8\} \rangle$ & $\frac32$ & $\frac32$ & $0$ & $0$ & $0$ & $\frac32$ & $\frac32$ & $0$ \\
$\langle \delta^{33} f^{8ge} f^{8he}\{T^g,T^h\} \rangle$ & $\frac94$ & $\frac94$ & $\frac32$ & $\frac32$ & $\frac32$ & $\frac94$ & $\frac94$ & $\frac92$ \\
$\langle \delta^{33} d^{8e8} T^e \rangle$ & $-\frac12$ & $-\frac12$ & $0$ & $0$ & $0$ & $\frac12$ & $\frac12$ & $0$ \\
$\langle \delta^{33} \delta^{88} \rangle$ & $1$ & $1$ & $1$ & $1$ & $1$ & $1$ & $1$ & $1$ \\
$\langle \{T^8,\{J^3,G^{38}\}\} \rangle$ & $\frac14$ & $\frac14$ & $0$ & $0$ & $0$ & $\frac34$ & $\frac34$ & $0$ \\
%$\langle \{T^8,\{J^3,G^{38}\}\} \rangle$ & $\frac14$ & $\frac14$ & $0$ & $0$ & $0$ & $\frac34$ & $\frac34$ & $0$ \\
$\langle \delta^{33} \{T^8,\{J^r,G^{r8}\}\} \rangle$ & $\frac34$ & $\frac34$ & $0$ & $0$ & $0$ & $\frac94$ & $\frac94$ & $0$ \\
%$\langle \delta^{33} \{T^8,\{J^r,G^{r8}\}\} \rangle$ & $\frac34$ & $\frac34$ & $0$ & $0$ & $0$ & $\frac94$ & $\frac94$ & $0$ \\
\end{tabular}
\end{ruledtabular}
\end{table*}

\begin{table*}
\caption{\label{t:bb2}Matrix elements of baryon operators corresponding to quadrupole moments of decuplet baryons for $N_c=3$: broken $SU(3)$ case.}
\begin{ruledtabular}
\begin{tabular}{lcccccccccc}
 & $\Delta^{++}$ & $\Delta^+$ & $\Delta^0$ & $\Delta^-$ & ${\Sigma^*}^+$ & ${\Sigma^*}^0$ & ${\Sigma^*}^-$ & ${\Xi^*}^-$ & ${\Xi^*}^0$ & $\Omega^-$ \\
\hline
$\langle \delta^{38} \{J^3,J^3\} \rangle$ & $0$ & $0$ & $0$ & $0$ & $0$ & $0$ & $0$ & $0$ & $0$ & $0$ \\
$\langle \delta^{33} \delta^{38} J^2 \rangle$ & $0$ & $0$ & $0$ & $0$ & $0$ & $0$ & $0$ & $0$ & $0$ & $0$ \\
$\langle d^{3e8} \{J^3,G^{3e}\} \rangle$ & $\frac{3\sqrt{3}}{4}$ & $\frac{\sqrt{3}}{4}$ & $-\frac{\sqrt{3}}{4}$ & $-\frac{3\sqrt{3}}{4}$ & $\frac{\sqrt{3}}{2}$ & $0$ & $-\frac{\sqrt{3}}{2}$ & $-\frac{\sqrt{3}}{4}$ & $\frac{\sqrt{3}}{4}$ & $0$ \\
$\langle \delta^{33} d^{3e8} \{J^r,G^{re}\} \rangle$ & $\frac{5\sqrt{3}}{4}$ & $\frac{5}{4\sqrt{3}}$ & $-\frac{5}{4\sqrt{3}}$ & $-\frac{5\sqrt{3}}{4}$ & $\frac{5}{2\sqrt{3}}$ & $0$ & $-\frac{5}{2\sqrt{3}}$ & $-\frac{5}{4\sqrt{3}}$ & $\frac{5}{4\sqrt{3}}$ & $0$ \\
$\langle d^{3e8} \{T^e,\{J^3,J^3\}\} \rangle$ & $\frac{9\sqrt{3}}{2}$ & $\frac{3\sqrt{3}}{2}$ & $-\frac{3\sqrt{3}}{2}$ & $-\frac{9\sqrt{3}}{2}$ & $3\sqrt{3}$ & $0$ & $-3\sqrt{3}$ & $-\frac{3\sqrt{3}}{2}$ & $\frac{3\sqrt{3}}{2}$ & $0$ \\
$\langle \delta^{33} d^{3e8} \{J^2,T^e\} \rangle$ & $\frac{15\sqrt{3}}{4}$ & $\frac{5\sqrt{3}}{4}$ & $-\frac{5\sqrt{3}}{4}$ & $-\frac{15\sqrt{3}}{4}$ & $\frac{5\sqrt{3}}{2}$ & $0$ & $-\frac{5\sqrt{3}}{2}$ & $-\frac{5\sqrt{3}}{4}$ & $\frac{5\sqrt{3}}{4}$ & $0$ \\
$\langle \{G^{33},G^{38}\} \rangle$ & $\frac{3\sqrt{3}}{8}$ & $\frac{\sqrt{3}}{8}$ & $-\frac{\sqrt{3}}{8}$ & $-\frac{3\sqrt{3}}{8}$ & $0$ & $0$ & $0$ & $\frac{\sqrt{3}}{8}$ & $-\frac{\sqrt{3}}{8}$ & $0$ \\
$\langle \delta^{33} \{T^3,T^8\} \rangle$ & $\frac{3\sqrt{3}}{2}$ & $\frac{\sqrt{3}}{2}$ & $-\frac{\sqrt{3}}{2}$ & $-\frac{3\sqrt{3}}{2}$ & $0$ & $0$ & $0$ & $\frac{\sqrt{3}}{2}$ & $-\frac{\sqrt{3}}{2}$ & $0$ \\
$\langle \delta^{33} f^{3ge} f^{8he}\{T^g,T^h\} \rangle$ & $\frac{3\sqrt{3}}{4}$ & $\frac{\sqrt{3}}{4}$ & $-\frac{\sqrt{3}}{4}$ & $-\frac{3\sqrt{3}}{4}$ & $\frac{3\sqrt{3}}{2}$ & $0$ & $-\frac{3\sqrt{3}}{2}$ & $-\frac{5\sqrt{3}}{4}$ & $\frac{5\sqrt{3}}{4}$ & $0$ \\
$\langle \delta^{33} d^{3e8} T^e \rangle$ & $\frac{\sqrt{3}}{2}$ & $\frac{1}{2\sqrt{3}}$ & $-\frac{1}{2\sqrt{3}}$ & $-\frac{\sqrt{3}}{2}$ & $\frac{1}{\sqrt{3}}$ & $0$ & $-\frac{1}{\sqrt{3}}$ & $-\frac{1}{2\sqrt{3}}$ & $\frac{1}{2\sqrt{3}}$ & $0$ \\
$\langle \delta^{33} \delta^{38} \rangle$ & $0$ & $0$ & $0$ & $0$ & $0$ & $0$ & $0$ & $0$ & $0$ & $0$ \\
$\langle \{T^3,\{J^3,G^{38}\}\} \rangle$ & $\frac{9\sqrt{3}}{4}$ & $\frac{3\sqrt{3}}{4}$ & $-\frac{3\sqrt{3}}{4}$ & $-\frac{9\sqrt{3}}{4}$ & $0$ & $0$ & $0$ & $\frac{3\sqrt{3}}{4}$ & $-\frac{3\sqrt{3}}{4}$ & $0$ \\
$\langle \{T^8,\{J^3,G^{33}\}\} \rangle$ & $\frac{9\sqrt{3}}{4}$ & $\frac{3\sqrt{3}}{4}$ & $-\frac{3\sqrt{3}}{4}$ & $-\frac{9\sqrt{3}}{4}$ & $0$ & $0$ & $0$ & $\frac{3\sqrt{3}}{4}$ & $-\frac{3\sqrt{3}}{4}$ & $0$ \\
$\langle \delta^{33} \{T^3,\{J^r,G^{r8}\}\} \rangle$ & $\frac{15\sqrt{3}}{4}$ & $\frac{5\sqrt{3}}{4}$ & $-\frac{5\sqrt{3}}{4}$ & $-\frac{15\sqrt{3}}{4}$ & $0$ & $0$ & $0$ & $\frac{5\sqrt{3}}{4}$ & $-\frac{5\sqrt{3}}{4}$ & $0$ \\
$\langle \delta^{33} \{T^8,\{J^r,G^{r3}\}\} \rangle$ & $\frac{15\sqrt{3}}{4}$ & $\frac{5\sqrt{3}}{4}$ & $-\frac{5\sqrt{3}}{4}$ & $-\frac{15\sqrt{3}}{4}$ & $0$ & $0$ & $0$ & $\frac{5\sqrt{3}}{4}$ & $-\frac{5\sqrt{3}}{4}$ & $0$ \\
$\langle \delta^{88} \{J^3,J^3\} \rangle$ & $\frac92$ & $\frac92$ & $\frac92$ & $\frac92$ & $\frac92$ & $\frac92$ & $\frac92$ & $\frac92$ & $\frac92$ & $\frac92$ \\
$\langle \delta^{33} \delta^{88} J^2 \rangle$ & $\frac{15}{4}$ & $\frac{15}{4}$ & $\frac{15}{4}$ & $\frac{15}{4}$ & $\frac{15}{4}$ & $\frac{15}{4}$ & $\frac{15}{4}$ & $\frac{15}{4}$ & $\frac{15}{4}$ & $\frac{15}{4}$ \\
$\langle d^{8e8} \{J^3,G^{3e}\} \rangle$ & $-\frac34$ & $-\frac34$ & $-\frac34$ & $-\frac34$ & $0$ & $0$ & $0$ & $\frac34$ & $\frac34$ & $\frac32$ \\
$\langle \delta^{33} d^{8e8} \{J^r,G^{re}\} \rangle$ & $-\frac54$ & $-\frac54$ & $-\frac54$ & $-\frac54$ & $0$ & $0$ & $0$ & $\frac54$ & $\frac54$ & $\frac52$ \\
$\langle d^{8e8} \{T^e,\{J^3,J^3\}\} \rangle$ & $-\frac92$ & $-\frac92$ & $-\frac92$ & $-\frac92$ & $0$ & $0$ & $0$ & $\frac92$ & $\frac92$ & $9$ \\
$\langle \delta^{33} d^{8e8} \{J^2,T^e\} \rangle$ & $-\frac{15}{4}$ & $-\frac{15}{4}$ & $-\frac{15}{4}$ & $-\frac{15}{4}$ & $0$ & $0$ & $0$ & $\frac{15}{4}$ & $\frac{15}{4}$ & $\frac{15}{2}$ \\
$\langle \{G^{38},G^{38}\} \rangle$ & $\frac38$ & $\frac38$ & $\frac38$ & $\frac38$ & $0$ & $0$ & $0$ & $\frac38$ & $\frac38$ & $\frac32$ \\
$\langle \delta^{33} \{T^8,T^8\} \rangle$ & $\frac32$ & $\frac32$ & $\frac32$ & $\frac32$ & $0$ & $0$ & $0$ & $\frac32$ & $\frac32$ & $6$ \\
$\langle \delta^{33} f^{8ge} f^{8he}\{T^g,T^h\} \rangle$ & $\frac94$ & $\frac94$ & $\frac94$ & $\frac94$ & $6$ & $6$ & $6$ & $\frac{27}{4}$ & $\frac{27}{4}$ & $\frac92$ \\
$\langle \delta^{33} d^{8e8} T^e \rangle$ & $-\frac12$ & $-\frac12$ & $-\frac12$ & $-\frac12$ & $0$ & $0$ & $0$ & $\frac12$ & $\frac12$ & $1$ \\
$\langle \delta^{33} \delta^{88} \rangle$ & $1$ & $1$ & $1$ & $1$ & $1$ & $1$ & $1$ & $1$ & $1$ & $1$ \\
$\langle \{T^8,\{J^3,G^{38}\}\} \rangle$ & $\frac94$ & $\frac94$ & $\frac94$ & $\frac94$ & $0$ & $0$ & $0$ & $\frac94$ & $\frac94$ & $9$ \\
%$\langle \{T^8,\{J^3,G^{38}\}\} \rangle$ & $\frac94$ & $\frac94$ & $\frac94$ & $\frac94$ & $0$ & $0$ & $0$ & $\frac94$ & $\frac94$ & $9$ \\
$\langle \delta^{33} \{T^8,\{J^r,G^{r8}\}\} \rangle$ & $\frac{15}{4}$ & $\frac{15}{4}$ & $\frac{15}{4}$ & $\frac{15}{4}$ & $0$ & $0$ & $0$ & $\frac{15}{4}$ & $\frac{15}{4}$ & $15$ \\
%$\langle \delta^{33} \{T^8,\{J^r,G^{r8}\}\} \rangle$ & $\frac{15}{4}$ & $\frac{15}{4}$ & $\frac{15}{4}$ & $\frac{15}{4}$ & $0$ & $0$ & $0$ & $\frac{15}{4}$ & $\frac{15}{4}$ & $15$ \\
\end{tabular}
\end{ruledtabular}
\end{table*}

\begin{table*}
\caption{\label{t:cc2}Matrix elements of baryon operators corresponding to transition quadrupole moments for $N_c=3$: broken $SU(3)$ case.}
\begin{ruledtabular}
\begin{tabular}{lcccccccc}
 &$\Delta^+p$ & $\Delta^0n$ & ${\Sigma^*}^0\Lambda$ & ${\Sigma^*}^0\Sigma^0$ & ${\Sigma^*}^+\Sigma^+$ & ${\Sigma^*}^-\Sigma^-$ & ${\Xi^*}^0\Xi^0$ & ${\Xi^*}^-\Xi^-$ \\
\hline
$\langle \delta^{38} \{J^3,J^3\} \rangle$ & $0$ & $0$ & $0$ & $0$ & $0$ & $0$ & $0$ & $0$ \\
$\langle \delta^{33} \delta^{38} J^2 \rangle$ & $0$ & $0$ & $0$ & $0$ & $0$ & $0$ & $0$ & $0$ \\
$\langle d^{3e8} \{J^3,G^{3e}\} \rangle$ & $\frac13\sqrt{\frac23}$ & $\frac13\sqrt{\frac23}$ & $\frac{1}{3\sqrt{2}}$ & $0$ & $\frac{1}{3\sqrt{6}}$ & $-\frac{1}{3\sqrt{6}}$ & $\frac{1}{3\sqrt{6}}$ & $-\frac{1}{3\sqrt{6}}$ \\
$\langle \delta^{33} d^{3e8} \{J^r,G^{re}\} \rangle$ & $0$ & $0$ & $0$ & $0$ & $0$ & $0$ & $0$ & $0$ \\
$\langle d^{3e8} \{T^e,\{J^3,J^3\}\} \rangle$ & $0$ & $0$ & $0$ & $0$ & $0$ & $0$ & $0$ & $0$ \\
$\langle \delta^{33} d^{3e8} \{J^2,T^e\} \rangle$ & $0$ & $0$ & $0$ & $0$ & $0$ & $0$ & $0$ & $0$ \\
$\langle \{G^{33},G^{38}\} \rangle$ & $\frac{1}{3\sqrt{6}}$ & $\frac{1}{3\sqrt{6}}$ & $-\frac{1}{3\sqrt{2}}$ & $0$ & $\frac13\sqrt{\frac23}$ & $-\frac13\sqrt{\frac23}$ & $-\frac{1}{3\sqrt{6}}$ & $\frac{1}{3 \sqrt{6}}$ \\
$\langle \delta^{33} \{T^3,T^8\} \rangle$ & $0$ & $0$ & $0$ & $0$ & $0$ & $0$ & $0$ & $0$ \\
$\langle \delta^{33} f^{3ge} f^{8he}\{T^g,T^h\} \rangle$ & $0$ & $0$ & $0$ & $0$ & $0$ & $0$ & $0$ & $0$ \\
$\langle \delta^{33} d^{3e8} T^e \rangle$ & $0$ & $0$ & $0$ & $0$ & $0$ & $0$ & $0$ & $0$ \\
$\langle \delta^{33} \delta^{38} \rangle$ & $0$ & $0$ & $0$ & $0$ & $0$ & $0$ & $0$ & $0$ \\
$\langle \{T^3,\{J^3,G^{38}\}\} \rangle$ & $0$ & $0$ & $0$ & $0$ & $\sqrt{\frac{2}{3}}$ & $-\sqrt{\frac{2}{3}}$ & $\frac{1}{\sqrt{6}}$ & $-\frac{1}{\sqrt{6}}$ \\
$\langle \{T^8,\{J^3,G^{33}\}\} \rangle$ & $\sqrt{\frac{2}{3}}$ & $\sqrt{\frac{2}{3}}$ & $0$ & $0$ & $0$ & $0$ & $-\frac{1}{\sqrt{6}}$ & $\frac{1}{\sqrt{6}}$ \\
$\langle \delta^{33} \{T^3,\{J^r,G^{r8}\}\} \rangle$ & $0$ & $0$ & $0$ & $0$ & $0$ & $0$ & $0$ & $0$ \\
$\langle \delta^{33} \{T^8,\{J^r,G^{r3}\}\} \rangle$ & $0$ & $0$ & $0$ & $0$ & $0$ & $0$ & $0$ & $0$ \\
$\langle \delta^{88} \{J^3,J^3\} \rangle$ & $0$ & $0$ & $0$ & $0$ & $0$ & $0$ & $0$ & $0$ \\
$\langle \delta^{33} \delta^{88} J^2 \rangle$ & $0$ & $0$ & $0$ & $0$ & $0$ & $0$ & $0$ & $0$ \\
$\langle d^{8e8} \{J^3,G^{3e}\} \rangle$ & $0$ & $0$ & $0$ & $-\frac{1}{3 \sqrt{2}}$ & $-\frac{1}{3 \sqrt{2}}$ & $-\frac{1}{3 \sqrt{2}}$ & $-\frac{1}{3 \sqrt{2}}$ & $-\frac{1}{3 \sqrt{2}}$ \\
$\langle \delta^{33} d^{8e8} \{J^r,G^{re}\} \rangle$ & $0$ & $0$ & $0$ & $0$ & $0$ & $0$ & $0$ & $0$ \\
$\langle d^{8e8} \{T^e,\{J^3,J^3\}\} \rangle$ & $0$ & $0$ & $0$ & $0$ & $0$ & $0$ & $0$ & $0$ \\
$\langle \delta^{33} d^{8e8} \{J^2,T^e\} \rangle$ & $0$ & $0$ & $0$ & $0$ & $0$ & $0$ & $0$ & $0$ \\
$\langle \{G^{38},G^{38}\} \rangle$ & $0$ & $0$ & $0$ & $\frac{1}{3\sqrt{2}}$ & $\frac{1}{3\sqrt{2}}$ & $\frac{1}{3\sqrt{2}}$ & $-\frac{\sqrt{2}}{3}$ & $-\frac{\sqrt{2}}{3}$ \\
$\langle \delta^{33} \{T^8,T^8\} \rangle$ & $0$ & $0$ & $0$ & $0$ & $0$ & $0$ & $0$ & $0$ \\
$\langle \delta^{33} f^{8ge} f^{8he}\{T^g,T^h\} \rangle$ & $0$ & $0$ & $0$ & $0$ & $0$ & $0$ & $0$ & $0$ \\
$\langle \delta^{33} d^{8e8} T^e \rangle$ & $0$ & $0$ & $0$ & $0$ & $0$ & $0$ & $0$ & $0$ \\
$\langle \delta^{33} \delta^{88} \rangle$ & $0$ & $0$ & $0$ & $0$ & $0$ & $0$ & $0$ & $0$ \\
$\langle \{T^8,\{J^3,G^{38}\}\} \rangle$ & $0$ & $0$ & $0$ & $0$ & $0$ & $0$ & $-\frac{1}{\sqrt{2}}$ & $-\frac{1}{\sqrt{2}}$ \\
%$\langle \{T^8,\{J^3,G^{38}\}\} \rangle$ & $0$ & $0$ & $0$ & $0$ & $0$ & $0$ & $\frac{1}{\sqrt{2}}$ & $-\frac{1}{\sqrt{2}}$ \\
$\langle \delta^{33} \{T^8,\{J^r,G^{r8}\}\} \rangle$ & $0$ & $0$ & $0$ & $0$ & $0$ & $0$ & $0$ & $0$ \\
%$\langle \delta^{33} \{T^8,\{J^r,G^{r8}\}\} \rangle$ & $0$ & $0$ & $0$ & $0$ & $0$ & $0$ & $0$ & $0$ \\
\end{tabular}
\end{ruledtabular}
\end{table*}

\section{\label{sec:full}A full expression for quadrupole moment}

The final expression for the quadrupole moment operator to linear order in SB is
\begin{equation}
\mathcal{Q}^{(ij)a} + \delta \mathcal{Q}^{(ij)a},
\end{equation}
where $\mathcal{Q}^{(ij)a}$ is the operator whose matrix elements yield the $SU(3)$ symmetric values; it is given in Eq.~(\ref{eq:qsym}). In turn, $\delta \mathcal{Q}^{(ij)a}$ includes all the operators due to first-order SB and its $1/N_c$ expansion reads
\begin{eqnarray}
\delta \mathcal{Q}^{(ij)a} & = & \sum_{n=2,4}^{N_c-1} \frac{1}{N_c^{n-1}} c_{n,\mathbf{1}} \mathcal{O}_{n,\mathbf{1}}^{(ij)a} + \sum_{n=2}^{N_c} \frac{1}{N_c^{n-1}} c_{n,\mathbf{8}} \mathcal{O}_{n,\mathbf{8}}^{(ij)a} + \sum_{n=2}^{N_c} \frac{1}{N_c^{n-1}} c_{n,\mathbf{27}} \mathcal{O}_{n,\mathbf{27}}^{(ij)a} + \sum_{n=4,6}^{N_c-1} \frac{1}{N_c^n} \tilde{c}_{n,\mathbf{27}} \tilde{\mathcal{O}}_{n,\mathbf{27}}^{(ij)a} \nonumber \\
&  & \mbox{} + \sum_{n=3,5}^{N_c} \frac{1}{N_c^{n-1}} c_{n,\mathbf{10}+\overline{\mathbf{10}}} \mathcal{O}_{n,\mathbf{10}+\overline{\mathbf{10}}}^{(ij)a}. \label{eq:qsbfull}
\end{eqnarray}

At the physical value $N_c=3$, the series is truncated as
\begin{eqnarray}
\delta \mathcal{Q}^{(ij)a} & = & \frac{1}{N_c} c_{2,\mathbf{1}} \mathcal{O}_{2,\mathbf{1}}^{(ij)a} + \frac{1}{N_c} c_{2,\mathbf{8}} \mathcal{O}_{2,\mathbf{8}}^{(ij)a} + \frac{1}{N_c^2} c_{3,\mathbf{8}} \mathcal{O}_{3,\mathbf{8}}^{(ij)a} + \frac{1}{N_c} c_{2,\mathbf{27}} \mathcal{O}_{2,\mathbf{27}}^{(ij)a} + \frac{1}{N_c^2} c_{3,\mathbf{27}} \mathcal{O}_{3,\mathbf{27}}^{(ij)a} \nonumber \\
&  & \mbox{} + \frac{1}{N_c^2} c_{3,\mathbf{10}+\overline{\mathbf{10}}} \mathcal{O}_{3,\mathbf{10}+\overline{\mathbf{10}}}^{(ij)a}, \label{eq:qsb}
\end{eqnarray}
where the operators $\mathcal{O}_{2,\mathbf{1}}^{(ij)a}$, $\mathcal{O}_{2,\mathbf{8}}^{(ij)a}$, $\mathcal{O}_{3,\mathbf{8}}^{(ij)a}$, $\mathcal{O}_{2,\mathbf{27}}^{(ij)a}$, $\mathcal{O}_{3,\mathbf{27}}^{(ij)a}$, and $\mathcal{O}_{3,\mathbf{10}+\overline{\mathbf{10}}}^{(ij)a}$ are given in Eqs.~(\ref{eq:212}), (\ref{eq:282}), (\ref{eq:283}), (\ref{eq:2272}), (\ref{eq:2273}), (\ref{eq:2103}), respectively. The operator coefficient $c_{n,\mathbf{rep}}$ accompany the $n$-body operator belonging to the flavor representation $\mathbf{rep}$. The matrix elements of the operators in the expansion (\ref{eq:qsb}) are listed in Tables \ref{t:aa2}, \ref{t:bb2}, and \ref{t:cc2} for the sake of completeness.

Thus, the complete quadrupole moments of decuplet baryons [$SU(3)$ symmetric value plus first-order SB effects] for $N_c=3$ are given by
\begin{subequations}
\label{eq:qdecc}
\begin{equation}
\mathcal{Q}_{\Delta^{++}} = \frac49 q_{\Delta^{++}} (k_2+k_3) + \frac{2}{3\sqrt{3}} c_{2,\mathbf{1}} + \frac{4}{9\sqrt{3}}(c_{2,\mathbf{8}}+c_{3,\mathbf{8}}) + \frac{3}{10\sqrt{3}} \left[ c_{2,\mathbf{27}} + \frac43 c_{3,\mathbf{27}} \right],
\end{equation}
\begin{equation}
\mathcal{Q}_{\Delta^+} = \frac49 q_{\Delta^+} (k_2+k_3) + \frac{2}{3\sqrt{3}} c_{2,\mathbf{1}} + \frac{1}{6\sqrt{3}} \left[ c_{2,\mathbf{27}} + \frac43 c_{3,\mathbf{27}} \right],
\end{equation}
\begin{equation}
\mathcal{Q}_{\Delta^0} = \frac{2}{3\sqrt{3}} c_{2,\mathbf{1}} - \frac{4}{9\sqrt{3}} (c_{2,\mathbf{8}} + c_{3,\mathbf{8}}) + \frac{1}{30\sqrt{3}} \left[ c_{2,\mathbf{27}} + \frac43c_{3,\mathbf{27}} \right], \label{eq:delta0}
\end{equation}
\begin{equation}
\mathcal{Q}_{\Delta^-} = \frac49 q_{\Delta^-} (k_2+k_3) + \frac{2}{3\sqrt{3}} c_{2,\mathbf{1}} - \frac{8}{9\sqrt{3}} (c_{2,\mathbf{8}} + c_{3,\mathbf{8}}) - \frac{1}{10\sqrt{3}} \left[ c_{2,\mathbf{27}} + \frac43c_{3,\mathbf{27}} \right],
\end{equation}
\begin{equation}
\mathcal{Q}_{{\Sigma^*}^+} = \frac49 q_{{\Sigma^*}^+} (k_2+k_3) + \frac{2}{3\sqrt{3}} c_{2,\mathbf{1}} + \frac{4}{9\sqrt{3}} (c_{2,\mathbf{8}} + c_{3,\mathbf{8}}) - \frac{11}{30\sqrt{3}} \left[ c_{2,\mathbf{27}} + \frac43c_{3,\mathbf{27}} \right],
\end{equation}
\begin{equation}
\mathcal{Q}_{{\Sigma^*}^0} = \frac{2}{3\sqrt{3}} c_{2,\mathbf{1}} - \frac{5}{30\sqrt{3}} \left[ c_{2,\mathbf{27}} + \frac43c_{3,\mathbf{27}} \right],
\end{equation}
\begin{equation}
\mathcal{Q}_{{\Sigma^*}^-} = \frac49 q_{{\Sigma^*}^-} (k_2+k_3) + \frac{2}{3\sqrt{3}} c_{2,\mathbf{1}} - \frac{4}{9\sqrt{3}} (c_{2,\mathbf{8}} + c_{3,\mathbf{8}}) + \frac{1}{30\sqrt{3}} \left[ c_{2,\mathbf{27}} + \frac43c_{3,\mathbf{27}} \right],
\end{equation}
\begin{equation}
\mathcal{Q}_{{\Xi^*}^-} = \frac49 q_{{\Xi^*}^-} (k_2+k_3) + \frac{2}{3\sqrt{3}} c_{2,\mathbf{1}} + \frac{5}{30\sqrt{3}} \left[ c_{2,\mathbf{27}} + \frac43c_{3,\mathbf{27}} \right],
\end{equation}
\begin{equation}
\mathcal{Q}_{{\Xi^*}^0} = \frac{2}{3\sqrt{3}} c_{2,\mathbf{1}} + \frac{4}{9\sqrt{3}} (c_{2,\mathbf{8}} + c_{3,\mathbf{8}}) - \frac{11}{30\sqrt{3}} \left[ c_{2,\mathbf{27}} + \frac43c_{3,\mathbf{27}} \right],
\end{equation}
and
\begin{equation}
\mathcal{Q}_{\Omega^-} = \frac49 q_{\Omega^-} (k_2+k_3) + \frac{2}{3\sqrt{3}} c_{2,\mathbf{1}} + \frac{4}{9\sqrt{3}} (c_{2,\mathbf{8}} + c_{3,\mathbf{8}}) + \frac{9}{30\sqrt{3}} \left[ c_{2,\mathbf{27}} + \frac43c_{3,\mathbf{27}} \right].
\end{equation}
\end{subequations}
Notice that SB effects induce nonvanishing contributions to the quadrupole moments of neutral decuplet baryons. Notice also that there is no contribution from the flavor $\mathbf{10}+\overline{\mathbf{10}}$ representation. 

For the transition quadrupole moments the expressions are
\begin{subequations}
\label{eq:qdecoctc}
\begin{equation}
\mathcal{Q}_{\Delta^+p} = \frac{2\sqrt{2}}{9} k_2 + \frac29 \sqrt{\frac23} c_{2,\mathbf{8}} + \frac{1}{15}
\sqrt{\frac23} \left[ c_{2,\mathbf{27}} + \frac13 c_{3,\mathbf{27}} \right] - \frac19 \sqrt{\frac23} c_{3,\mathbf{10}+\overline{\mathbf{10}}},
\end{equation}
\begin{equation}
\mathcal{Q}_{\Delta^0n} = \frac{2\sqrt{2}}{9} k_2 + \frac29 \sqrt{\frac23} c_{2,\mathbf{8}} + \frac{1}{15} \sqrt{\frac23} \left[ c_{2,\mathbf{27}} + \frac13 c_{3,\mathbf{27}} \right] - \frac19 \sqrt{\frac23} c_{3,\mathbf{10}+\overline{\mathbf{10}}},
\end{equation}
\begin{equation}
\mathcal{Q}_{{\Sigma^*}^0\Lambda} = \frac13 \sqrt{\frac23} k_2 + \frac{\sqrt{2}}{9} c_{2,\mathbf{8}} - \frac{2\sqrt{2}}{15} \left[ c_{2,\mathbf{27}} + \frac13 c_{3,\mathbf{27}} \right],
\end{equation}
\begin{equation}
\mathcal{Q}_{{\Sigma^*}^0\Sigma^0} = \frac{\sqrt{2}}{9} k_2 - \frac19 \sqrt{\frac23} c_{2,\mathbf{8}} + \frac{2}{15}
\sqrt{\frac23} \left[ c_{2,\mathbf{27}} + \frac13c_{3,\mathbf{27}} \right],
\end{equation}
\begin{equation}
\mathcal{Q}_{{\Sigma^*}^+\Sigma^+} = \frac{2\sqrt{2}}{9} k_2 + \frac13 \sqrt{\frac23} \left[ c_{2,\mathbf{27}} +\frac13 c_{3,\mathbf{27}} \right] + \frac19 \sqrt{\frac23} c_{3,\mathbf{10}+\overline{\mathbf{10}}},
\end{equation}
\begin{equation}
\mathcal{Q}_{{\Sigma^*}^-\Sigma^-} = -\frac29 \sqrt{\frac23} c_{2,\mathbf{8}} - \frac{1}{15} \sqrt{\frac23} \left[ c_{2,\mathbf{27}} + \frac13 c_{3,\mathbf{27}} \right] - \frac19 \sqrt{\frac23} c_{3,\mathbf{10}+\overline{\mathbf{10}}},
\end{equation}
\begin{equation}
\mathcal{Q}_{{\Xi^*}^0\Xi^0} = \frac{2\sqrt{2}}{9} k_2 - \frac13 \sqrt{\frac23} \left[ c_{2,\mathbf{27}} + \frac13 c_{3,\mathbf{27}} \right] + \frac19 \sqrt{\frac23} c_{3,\mathbf{10}+\overline{\mathbf{10}}},
\end{equation}
\begin{equation}
\mathcal{Q}_{{\Xi^*}^-\Xi^-} = -\frac29 \sqrt{\frac23} c_{2,\mathbf{8}} - \frac{1}{15} \sqrt{\frac23} \left[ c_{2,\mathbf{27}} + \frac13 c_{3,\mathbf{27}} \right] - \frac19 \sqrt{\frac23} c_{3,\mathbf{10}+\overline{\mathbf{10}}}.
\end{equation}
\end{subequations}
Notice that the flavor singlet representation does not contribute in this case but the flavor $\mathbf{10}+\overline{\mathbf{10}}$ one does.

The isospin relations listed in Sec.~\ref{sec:su3}, this time for the complete expressions for quadrupole moments are, for $I=3$ operators,
\begin{subequations}
\label{eq:isosb}
\begin{equation}
\mathcal{Q}_{\Delta^{++}} - 3 \mathcal{Q}_{\Delta^+} + 3 \mathcal{Q}_{\Delta^0} - \mathcal{Q}_{\Delta^-} = 0,
\end{equation}
and for $I=2$ operators,
\begin{equation}
\mathcal{Q}_{\Delta^{++}} - \mathcal{Q}_{\Delta^+} - \mathcal{Q}_{\Delta^0} + \mathcal{Q}_{\Delta^-} = 0,
\end{equation}
\begin{equation}
\mathcal{Q}_{{\Sigma^*}^+} - 2\mathcal{Q}_{{\Sigma^*}^0} + \mathcal{Q}_{{\Sigma^*}^-} = 0,
\end{equation}
\begin{equation}
\mathcal{Q}_{\Delta^+p} - \mathcal{Q}_{\Delta^0n} = 0,
\end{equation}
\begin{equation}
\mathcal{Q}_{{\Sigma^*}^+\Sigma^+} - 2\mathcal{Q}_{{\Sigma^*}^0\Sigma^0} + \mathcal{Q}_{{\Sigma^*}^-\Sigma^-} = 0,
\end{equation}
\end{subequations}
\textit{i.e.}, they are fulfilled \textit{in the presence of first-order SB}, which is a completely expected result.

Additional expressions fulfilled in the presence of first-order SB are
\begin{subequations}
\label{eq:otherr}
\begin{equation}
(\mathcal{Q}_{{\Sigma^*}^-} + \mathcal{Q}_{{\Sigma^*}^+}) - (\mathcal{Q}_{\Delta^0} + \mathcal{Q}_{{\Xi^*}^0}) = 0,
\end{equation}
\begin{equation}
\frac23 (\mathcal{Q}_{\Delta^{++}} - \mathcal{Q}_{\Omega^-}) - (\mathcal{Q}_{\Delta^+} - \mathcal{Q}_{{\Xi^*}^-}) = 0,
\end{equation}
\begin{equation}
\mathcal{Q}_{\Delta^+} - 2 \mathcal{Q}_{\Delta^0} + \mathcal{Q}_{\Delta^-} = 0,
\end{equation}
\begin{equation}
-2 \mathcal{Q}_{{\Sigma^*}^0} + \mathcal{Q}_{\Delta^0} + \mathcal{Q}_{{\Xi^*}^0} = 0,
\end{equation}
\begin{equation}
\mathcal{Q}_{{\Sigma^*}^-} + \mathcal{Q}_{\Delta^-} - 5 \mathcal{Q}_{{\Xi^*}^-} + 3 \mathcal{Q}_{\Omega^-} = 0,
\end{equation}
\begin{equation}
\mathcal{Q}_{{\Sigma^*}^-\Sigma^-} - \mathcal{Q}_{{\Xi^*}^-\Xi^-} = 0, \label{eq:ssmsm}
\end{equation}
\end{subequations}
Relation (\ref{eq:otherr}f) is remarkable because in the $SU(3)$ limit, $\mathcal{Q}_{{\Sigma^*}^-\Sigma^-}^{SU(3)} = \mathcal{Q}_{{\Xi^*}^-\Xi^-}^{SU(3)} = 0$, and still the degeneracy between these two quantities is not lifted by first-order SB effects. In consequence, if $\epsilon \sim m_s$ is a (dimensionless measure) of $SU(3)$ breaking, corrections to Eqs.~(\ref{eq:otherr}) should arise are order $\epsilon^2$.

Flavor SB is evaluated in Ref.~\cite{bh2} by replacing the spin-spin terms in the expressions for the quadrupole moments with a {\it quadratic} quark mass dependence as obtained from a one-gluon exchange interaction between the quarks. SB is then characterized by the ratio $r=m_u/m_s$ of $u$ and $s$ quark masses. The counterparts of relations (\ref{eq:bhrel}) with SB effects are also found to vanish in Ref.~\cite{bh2}. However, in the formalism presented here, they are now given by
\begin{subequations}
\begin{equation}
\mathcal{Q}_{\Delta^-} + \mathcal{Q}_{\Delta^+} = \frac{4}{3\sqrt{3}} c_{2,\mathbf{1}} - \frac{8}{9\sqrt{3}} c_{2,\mathbf{8}} - \frac{8}{9\sqrt{3}} c_{3,\mathbf{8}} + \frac{1}{15 \sqrt{3}} c_{2,\mathbf{27}} + \frac{4}{45\sqrt{3}} c_{3,\mathbf{27}},
\end{equation}
\begin{equation}
2 \mathcal{Q}_{\Delta^-}+\mathcal{Q}_{\Delta^{++}} = \frac{2}{\sqrt{3}} c_{2,\mathbf{1}} - \frac{4}{3\sqrt{3}} c_{2,\mathbf{8}} - \frac{4}{3\sqrt{3}} c_{3,\mathbf{8}} + \frac{1}{10\sqrt{3}} c_{2,\mathbf{27}} + \frac{2 }{15\sqrt{3}} c_{3,\mathbf{27}} ,
\end{equation}
\begin{equation}
3 (\mathcal{Q}_{{\Xi^*}^-} - \mathcal{Q}_{{\Sigma^*}^-}) - (\mathcal{Q}_{\Omega^-} - \mathcal{Q}_{\Delta^-}) = 0, \label{eq:equa}
\end{equation}
\begin{equation}
\mathcal{Q}_{\Delta^-} - \mathcal{Q}_{{\Sigma^*}^-} - \sqrt{2} \mathcal{Q}_{{\Sigma^*}^- \Sigma^-} = - \frac{4}{9\sqrt{3}} c_{3,\mathbf{8}}- \frac{2}{15\sqrt{3}} c_{3,\mathbf{27}} + \frac{2}{9\sqrt{3}} c_{3,\mathbf{10}+\overline{\mathbf{10}}},
\end{equation}
\begin{equation}
\mathcal{Q}_{\Delta^+} - \mathcal{Q}_{{\Sigma^*}^+} + \sqrt{2} \mathcal{Q}_{\Delta^+ p} - \sqrt{2} \mathcal{Q}_{{\Sigma^*}^+ \Sigma^+} = - \frac{4}{9\sqrt{3}} c_{3,\mathbf{8}} + \frac{8}{15\sqrt{3}} c_{3,\mathbf{27}} - \frac{4}{9\sqrt{3}} c_{3,\mathbf{10}+\overline{\mathbf{10}}},
\end{equation}
\begin{equation}
\mathcal{Q}_{{\Sigma^*}^0} - \frac{1}{\sqrt{2}} \mathcal{Q}_{{\Sigma^*}^0 \Sigma^0} + \frac{1}{\sqrt{6}} \mathcal{Q}_{{\Sigma^*}^0 \Lambda} =
\frac{2}{3\sqrt{3}} c_{2,\mathbf{1}} + \frac{2}{9\sqrt{3}} c_{2,\mathbf{8}} - \frac{13}{30\sqrt{3}} c_{2,\mathbf{27}} - \frac{14}{45\sqrt{3}} c_{3,\mathbf{27}},
\end{equation}
\begin{equation}
\mathcal{Q}_{{\Sigma^*}^-} - \mathcal{Q}_{{\Xi^*}^-} - \frac{1}{\sqrt{2}} \mathcal{Q}_{{\Xi^*}^- \Xi^-} - \frac{1}{\sqrt{2}} \mathcal{Q}_{{\Sigma^*}^- \Sigma^-} = - \frac{4}{9\sqrt{3}} c_{3,\mathbf{8}} - \frac{2}{15\sqrt{3}} c_{3,\mathbf{27}} + \frac{2}{9\sqrt{3}} c_{3,\mathbf{10}+\overline{\mathbf{10}}},
\end{equation}
\begin{equation}
\mathcal{Q}_{{\Xi^*}^0} + \frac{1}{\sqrt{2}} \mathcal{Q}_{{\Xi^*}^0 \Xi^0} - \sqrt{\frac{2}{3}} \mathcal{Q}_{{\Sigma^*}^0 \Lambda} =
\frac{2}{3\sqrt{3}} c_{2,\mathbf{1}} + \frac{2}{9\sqrt{3}} c_{2,\mathbf{8}} + \frac{4}{9\sqrt{3}} c_{3,\mathbf{8}} - \frac{13}{30\sqrt{3}} c_{2,\mathbf{27}} - \frac{23}{45\sqrt{3}}  c_{3,\mathbf{27}} + \frac{1}{9\sqrt{3}} c_{3,\mathbf{10}+\overline{\mathbf{10}}}.
\end{equation}
\end{subequations}

Equation (\ref{eq:equa}) is the only agreement between the analysis of Ref.~\cite{bh2} and the present one, as far as SB effects in the quadrupole moments are concerned. A noticeable difference lies in the vanishing value of $\mathcal{Q}_{\Delta^0}$ in the presence of SB even to order $\mathcal{O}(r^3)$ found in that reference, compared to Eq.~(\ref{eq:delta0}), which attains in principle a nonzero value precisely due to first-order SB.

In the context of the $1/N_c$ expansion analysis of Ref.~\cite{bl2}, an additional relation is provided with $N_c$-independent coefficients that holds for all values of $N_c$ in all cases of SB analyzed there. The relation, corresponding to Eq.~(4.23) in that reference, reads
\begin{equation}
\mathcal{Q}_{{\Xi^*}^-} - \mathcal{Q}_{\Omega^-} - \sqrt{2} \mathcal{Q}_{{\Xi^*}^-\Xi^-} = 0.
\end{equation}

With the results presented here, the above relation actually reads
\begin{equation}
\mathcal{Q}_{{\Xi^*}^-} - \mathcal{Q}_{\Omega^-} - \sqrt{2} \mathcal{Q}_{{\Xi^*}^-\Xi^-} = - \frac{4}{9\sqrt{3}} c_{3,\mathbf{8}} - \frac{2}{15\sqrt{3}} c_{3,\mathbf{27}} + \frac{2}{9\sqrt{3}} c_{3,\mathbf{10}+\overline{\mathbf{10}}}. \label{eq:423}
\end{equation}
Hence, SB affects relation (\ref{eq:423}) \textit{only} at 3-body operator level and higher. Should these operators be removed from the series (\ref{eq:qsb}), the single-photon exchange ansatz prediction would be recovered. Equivalently, relation (\ref{eq:423}) vanishes when leading and subleading terms in $1/N_c$ are retained in the series (\ref{eq:qsb}). Beyond that point, it gets modifications.

\section{\label{sec:clo}Closing remarks}

The main aim of the present paper is to construct the $1/N_c$ expansion of the baryon operator whose matrix elements between $SU(6)$ baryon states yields the actual values of the \textit{spectroscopic} quadrupole moments. This operator has well-defined properties: It is a spin-2 and a flavor octet object, which means that it transforms as $(2,\mathbf{8})$ under $SU(2)\times SU(3)$. It is a symmetric and traceless tensor in the spin indices. And most importantly, it is even under time reversal.

The operator is first constructed under the assumption of an exact $SU(3)$ flavor symmetry; it is denoted by $\mathcal{Q}^{(ij)a}$. For the physical value $N_c=3$, $\mathcal{Q}^{(ij)a}$ is given by Eq.~(\ref{eq:1onq}). The effects of SB are accounted for to linear order through the operator $\delta \mathcal{Q}^{(ij)a}$, given by Eq.~(\ref{eq:qsb}), which comprises all the operators that fall into the flavor representations allowed by the tensor product of the quadrupole moment and the perturbation. These representations are $(2,\mathbf{1})$, $(2,\mathbf{8})$, $(2,\mathbf{8})$, $(2,\mathbf{10}+\overline{\mathbf{10}})$, and $(2,\mathbf{27})$. The $1/N_c$ expansions of the operators that satisfy the properties mentioned above are given in detailed. The matrix elements of the operator $\mathcal{Q}^{(ij)a}+\delta \mathcal{Q}^{(ij)a}$ yield the actual values of the quadrupole moments. They are listed in Eqs.~(\ref{eq:qdecc}) and (\ref{eq:qdecoctc}) for baryon decuplet and baryon decuplet-octet transitions. For baryon octet, the values are found to be zero, which is consistent with angular momentum selection rules. These expressions are given in terms of the free parameters of the theory. Retaining up to 3-body operators, there are two parameters ($k_2$ and $k_3$) for the case of exact flavor symmetry and six more introduced by SB ($c_{2,\mathbf{1}}$, $c_{2,\mathbf{8}}$, $c_{3,\mathbf{8}}$, $c_{2,\mathbf{27}}$, $c_{3,\mathbf{27}}$, and $c_{3,\mathbf{10}+\overline{\mathbf{10}}}$). All in all, there are eight undetermined parameters. Unfortunately, the experimental information \cite{part} is rather scarce, so at this time, it is not possible to perform a least-squares fit to compare theory and experiment and extract information on these parameters. Any other attempts of reducing the number of free parameters are fruitless. The only pieces of information known up to now are those corresponding to the transition $\Delta^+\to p$ \cite{part}. The difficulties in measuring quadrupole moments depend on many factors. For instance, except for $\Omega^-$, all of the decuplet baryons decay strongly, so couplings of the form decuplet-decuplet-$\gamma$ are available only through virtual processes, which are difficult to measure \cite{bl1}.

To overcome the lack of experimental information on quadrupole moments, some relations among them are provided instead. Apart from the isospin relations (\ref{eq:isosb}) that must be satisfied by quadrupole moments in the presence of SB, other relations are also provided, (\ref{eq:otherr}), which can be quite useful in the future, when additional experiments are envisaged. In the meantime, the predictions of the $1/N_c$ expansion are in accordance with expectations.

\begin{acknowledgments}
The authors are grateful to Consejo Nacional de Ciencia y Tecnolog{\'\i}a (Mexico) for partial support. R.F.-M.\ was also partially supported by Fondo de Apoyo a la Investigaci\'on (Universidad Aut\'onoma de San Luis Potos{\'\i}).
\end{acknowledgments}

\appendix*

\section{\label{app:reduc}Reduction of baryon operators}

Different $n$-body operators that satisfy the properties imposed to make up the $1/N_c$ expansion of $\mathcal{Q}^{(ij)a}$ can be constructed. However, some of them are linearly dependent and can be written in terms of the chosen linearly independent ones by using operator identities \cite{djm95}. In this section a list of operator reductions, as complete as possible, is presented to identify those operators which are not eligible in the $1/N_c$ expansion of $\mathcal{Q}^{(ij)a}$. Each one of the relations provided contains, on the left-hand side, the presumably dependent operator whereas, on the right-hand side, its equivalence in terms of the chosen operator basis.

\subsection{$2$-body operators}

\begin{equation}
\{G^{ie},G^{je}\} - \frac13 \delta^{ij} \{G^{re},G^{re}\} = \frac{N_f-1}{2N_f} \left[ \{J^i,J^j\} - \frac23 \delta^{ij} J^2 \right],
\end{equation}
\begin{equation}
d^{abc} (\{G^{ib},G^{jc}\} + \{G^{jb},G^{ic}\}) - \frac23 \delta^{ij} d^{abc} \{G^{rb},G^{rc}\} = \frac{N_f-2}{N_f} \left[ \{J^i,G^{ja}\}+\{J^j,G^{ia}\} - \frac23 \delta^{ij} \{J^r,G^{ra}\} \right].
\end{equation}

\subsection{$3$-body operators}

\begin{equation}
\{J^i,\{T^e,G^{je}\}\} + \{J^j,\{T^e,G^{ie}\}\} - \frac23 \delta^{ij} \{J^r,\{T^e,G^{re}\}\} = \frac{2(N_c+N_f)(N_f-1)}{N_f} \left[ \{J^i,J^j\} - \frac23 \delta^{ij} J^2 \right], \label{eq:jtg}
\end{equation}
\begin{equation}
\{T^a,\{G^{ie}, G^{je}\}\} - \frac13 \delta^{ij} \{T^a,\{G^{re}, G^{re}\}\} = \frac{N_f-1}{2N_f} \left[ \{T^a,\{J^i,J^j\}\} - \frac23 \delta^{ij} \{J^2,T^a\} \right],
\end{equation}
\begin{eqnarray}
&  & \{G^{ia},\{T^e, G^{je}\}\} + \{G^{ja},\{T^e,G^{ie}\}\} - \frac23 \delta^{ij} \{G^{ra},\{T^e,G^{re}\}\} = \nonumber \\
&  & \mbox{} \frac{(N_c + N_f)(N_f-1)}{N_f} \left[ \{J^i,G^{ja}\} + \{J^j,G^{ia}\} - \frac23 \delta^{ij} \{J^r,G^{ra}\} \right],
\end{eqnarray}
\begin{equation}
\{G^{ie},\{T^a,G^{je}\}\} + \{G^{je},\{T^a,G^{ie}\}\} - \frac23 \delta^{ij} \{G^{re},\{T^a,G^{re}\}\} = \frac{N_f-1}{N_f} \left[ \{T^a,\{J^i,J^j\}\} - \frac23 \delta^{ij} \{J^2,T^a\} \right],
\end{equation}
\begin{eqnarray}
&  & d^{abc} (\{J^i,\{T^b,G^{jc}\}\}+\{J^j,\{T^b,G^{ic}\}\}) - \frac23 \delta^{ij} d^{abc} \{J^r,\{T^b,G^{rc}\}\} = \frac{N_f-2}{N_f} \left[ \{T^a,\{J^i,J^j\}\} - \frac23 \delta^{ij} \{J^2,T^a\} \right] \nonumber \\
&  & \mbox{} + \frac{(N_c+N_f)(N_f-2)}{N_f} \left[ \{J^i,G^{ja}\}+\{J^j,G^{ia}\} - \frac23 \delta^{ij}\{J^r,G^{ra}\} \right].
\end{eqnarray}

\subsection{$4$-body operators}

\begin{eqnarray}
&  & d^{bcd} (\{G^{ia},\{G^{jb},\{G^{rc},G^{rd}\}\}\} + \{G^{ja},\{G^{ib},\{G^{rc},G^{rd}\}\}\}) - \frac23 \delta^{ij} d^{bcd} \{G^{ma},\{G^{mb},\{G^{rc},G^{rd}\}\}\} = \nonumber \\
&  & \frac{N_f-2}{8N_f}[5N_c(N_c+2N_f)+6N_f^2+2N_f-8] \left[ \{J^i,G^{ja}\} + \{J^j,G^{ia}\} - \frac23 \delta^{ij} \{J^r,G^{ra}\} \right] \nonumber \\
&  & \mbox{} - \frac{(N_f+4)(N_f-2)}{4N_f^2} \left[ \{J^2,\{J^i,G^{ja}\}\} + \{J^2,\{J^j,G^{ia}\}\} - \frac23 \delta^{ij} \{J^2,\{J^r,G^{ra}\}\} \right],
\end{eqnarray}
\begin{eqnarray}
&  & d^{abc} \{\{G^{ib},G^{jc}\},\{G^{re},G^{re}\}\} - \frac13 \delta^{ij} d^{abc} \{\{G^{mb},G^{mc}\},\{G^{re},G^{re}\}\} = \nonumber \\
&  & \mbox{} \frac38 \frac{N_c(N_c+2N_f)(N_f-2)}{N_f} \left[\{J^i,G^{ja}\} + \{J^j,G^{ia}\} - \frac23 \delta^{ij} \{J^r,G^{ra}\} \right] \nonumber \\
&  & \mbox{} - \frac{N_f^2-4}{4N_f^2} \left[\{J^2,\{J^i,G^{ja}\}\} + \{J^2,\{J^j,G^{ia}\}\} - \frac23 \delta^{ij} \{J^2,\{J^r,G^{ra}\}\} \right],
\end{eqnarray}
\begin{eqnarray}
&  & d^{abc} \{\{J^i,J^j\},\{T^b,T^c\}\} - \frac13 \delta^{ij} d^{abc} \{\{J^r,J^r\},\{T^b,T^c\}\} = \nonumber \\
&  & \mbox{} 2 \left[ \{\{J^i,J^j\},\{J^r,G^{ra}\}\} - \frac23 \delta^{ij} \{J^2,\{J^r,G^{ra}\}\} \right] + \frac{(N_c+N_f)(N_f-4)}{N_f} \left[ \{T^a,\{J^i,J^j\}\} - \frac23 \delta^{ij} \{J^2,T^a\} \right],
\end{eqnarray}
\begin{eqnarray}
&  & d^{abc} \{\{J^i,J^j\},\{G^{rb},G^{rc}\}\} - \frac23 \delta^{ij} d^{abc} \{J^2,\{G^{rb},G^{rc}\}\} = \nonumber \\
&  & \mbox{} -\frac{N_f+4}{2N_f} \left[ \{\{J^i,J^j\},\{J^r,G^{ra}\}\} - \frac23 \delta^{ij} \{J^2,\{J^r,G^{ra}\}\}\right] + \frac34 (N_c+N_f) \left[ \{T^a,\{J^i,J^j\}\} - \frac23 \delta^{ij} \{J^2,T^a\} \right],
\end{eqnarray}
\begin{eqnarray}
&  & \{\{J^i,G^{ja}\},\{G^{re},G^{re}\}\} + \{\{J^j,G^{ia}\},\{G^{re},G^{re}\}\} - \frac23 \delta^{ij} \{\{J^m,G^{ma}\},\{G^{re},G^{re}\}\} = \nonumber \\
&  & \mbox{} \frac34 N_c(N_c+2N_f) \left[ \{J^i,G^{ja}\} + \{J^j,G^{ia}\} - \frac23 \delta^{ij} \{J^r,G^{ra}\} \right] \nonumber \\
&  & \mbox{} - \frac{N_f+2}{2 N_f} \left[\{J^2,\{J^i,G^{ja}\}\} + \{J^2,\{J^j,G^{ia}\}\} - \frac23 \delta^{ij} \{J^2,\{J^r,G^{ra}\}\} \right],
\end{eqnarray}
\begin{equation}
\{\{J^r,G^{ra}\},\{G^{ie},G^{je}\}\} - \frac13 \delta^{ij} \{\{J^r,G^{ra}\},\{G^{me},G^{me}\}\} = \frac{N_f-1}{2N_f} \left[ \{\{J^i,J^j\},\{J^r,G^{ra}\}\} - \frac23 \delta^{ij} \{J^2,\{J^r,G^{ra}\}\} \right].
\end{equation}

\subsection{$5$-body operators}

\begin{eqnarray}
&  & d^{abc} \{\{J^i,J^j\},\{J^r,\{T^b,G^{rc}\}\}\} - \frac23 \delta^{ij} d^{abc} \{J^2,\{J^r,\{T^b,G^{rc}\}\}\} = \nonumber \\
&  & \mbox{} \frac{N_f-2}{N_f} \left[ \{J^2,\{T^a,\{J^i,J^j\}\}\} - \frac23 \delta^{ij} \{J^2,\{J^2,T^a\}\} \right] \nonumber \\
&  & \mbox{} + \frac{(N_c+N_f)(N_f-2)}{N_f} \left[ \{\{J^i,J^j\},\{J^r,G^{ra}\}\} - \frac23 \delta^{ij} \{J^2,\{J^r,G^{ra}\}\} \right],
\end{eqnarray}
\begin{eqnarray}
&  & \{T^a,\{\{G^{ic},G^{jc}\},\{G^{re},G^{re}\}\}\} - \frac13 \delta^{ij} \{T^a,\{\{G^{mc},G^{mc}\},\{G^{re},G^{re}\}\}\} = \nonumber \\
&  & \mbox{} - \frac14 \frac{(N_f+2)(N_f-1)}{N_f^2} \left[ \{J^2,\{T^a,\{J^i,J^j\}\}\} - \frac23 \delta^{ij}\{J^2,\{J^2,T^a\}\} \right] \nonumber \\
&  & \mbox{} + \frac38 \frac{N_c(N_c+2N_f)(N_f-1)}{N_f} \left[ \{T^a,\{J^i,J^j\}\} - \frac23 \delta^{ij} \{J^2,T^a\} \right].
\end{eqnarray}

\end{document}